\documentclass[a4paper,12pt,onecolumn,draftclsnofoot]{IEEEtran}
\usepackage{amssymb}
\usepackage{tabularx}
\usepackage{graphicx}
\usepackage{caption}
\usepackage{subcaption}
\usepackage[tbtags]{amsmath}
\usepackage{amsfonts}
\usepackage{mathrsfs}
\usepackage{multirow}
\usepackage{cite}
\usepackage{url}
\usepackage{hyperref}
\hypersetup{
  colorlinks=true,
  linkcolor=red,
  linktoc=all,
  citecolor=blue,
  bookmarksnumbered=true
}
\usepackage{units}

\usepackage[section]{placeins}

\IEEEoverridecommandlockouts

\begin{document}
\title{City-Scale Agent-Based Simulators for the Study of Non-Pharmaceutical Interventions in the Context of the COVID-19 Epidemic}
\author{
\IEEEauthorblockN{
IISc-TIFR COVID-19 City-Scale Simulation Team}
\thanks{
Authors in alphabetical order of last names:
Shubhada Agrawal\IEEEauthorrefmark{2},
Siddharth Bhandari\IEEEauthorrefmark{2},
Anirban Bhattacharjee\IEEEauthorrefmark{2},
Anand Deo\IEEEauthorrefmark{2},
Narendra M. Dixit\IEEEauthorrefmark{1},
Prahladh Harsha\IEEEauthorrefmark{2},
Sandeep Juneja\IEEEauthorrefmark{2},
Poonam Kesarwani\IEEEauthorrefmark{2},
Aditya Krishna Swamy\IEEEauthorrefmark{1},
Preetam Patil\IEEEauthorrefmark{1},
Nihesh Rathod\IEEEauthorrefmark{1},
Ramprasad Saptharishi\IEEEauthorrefmark{2},
Sharad Shriram\IEEEauthorrefmark{1},
Piyush Srivastava\IEEEauthorrefmark{2},
Rajesh Sundaresan\IEEEauthorrefmark{1},
Nidhin Koshy Vaidhiyan\IEEEauthorrefmark{1},
Sarath Yasodharan\IEEEauthorrefmark{1}}
\thanks{\textbf{Corresponding author: Rajesh Sundaresan,  rajeshs@iisc.ac.in}}\\
\thanks{PP, NR, SS, NKV, SY from IISc were supported by the IISc-Cisco Centre for Networked Intelligence, Indian Institute of Science. RSun was supported by the IISc-Cisco Centre for Networked Intelligence, the Robert Bosch Centre for Cyber-Physical Systems, and the Department of Electrical Communication Engineering, Indian Institute of Science.}
\thanks{TIFR co-authors acknowledge support of the Department of Atomic Energy, Government of India, under project no.~RTI4001.}
\thanks{PH acknowledges support from the Swarnajayanti Fellowship of DST.}
\thanks{RSap acknowledges support from the Ramanujan Fellowship of SERB.}
\thanks{PS acknowledges support from the Ramanujan Fellowship of SERB, from Adobe Systems Inc. via a gift to TIFR, and from the Infosys-Chandrasekharan virtual centre for Random Geometry supported by a grant from the Infosys Foundation.}
\thanks{The contents of this paper do not necessarily reflect the views of the funding agencies.}
\thanks{Source code available at:~\url{https://github.com/cni-iisc/epidemic-simulator/releases/tag/3.0}}
\IEEEauthorblockA{
\IEEEauthorrefmark{1}Indian Institute of Science, Bengaluru\\
\IEEEauthorrefmark{2}TIFR, Mumbai}\\
11 August 2020
}

\maketitle

\begin{abstract}
We highlight the usefulness of city-scale agent-based simulators in studying various non-pharmaceutical interventions to manage an evolving pandemic. We ground our studies in the context of the COVID-19 pandemic and demonstrate the power of the simulator via several exploratory case studies in two metropolises, Bengaluru and Mumbai. Such tools become common-place in any city administration's tool kit in our march towards digital health.
\end{abstract}

\section{Introduction}
\begin{figure*}[t]
\centering
\includegraphics[scale=0.5]{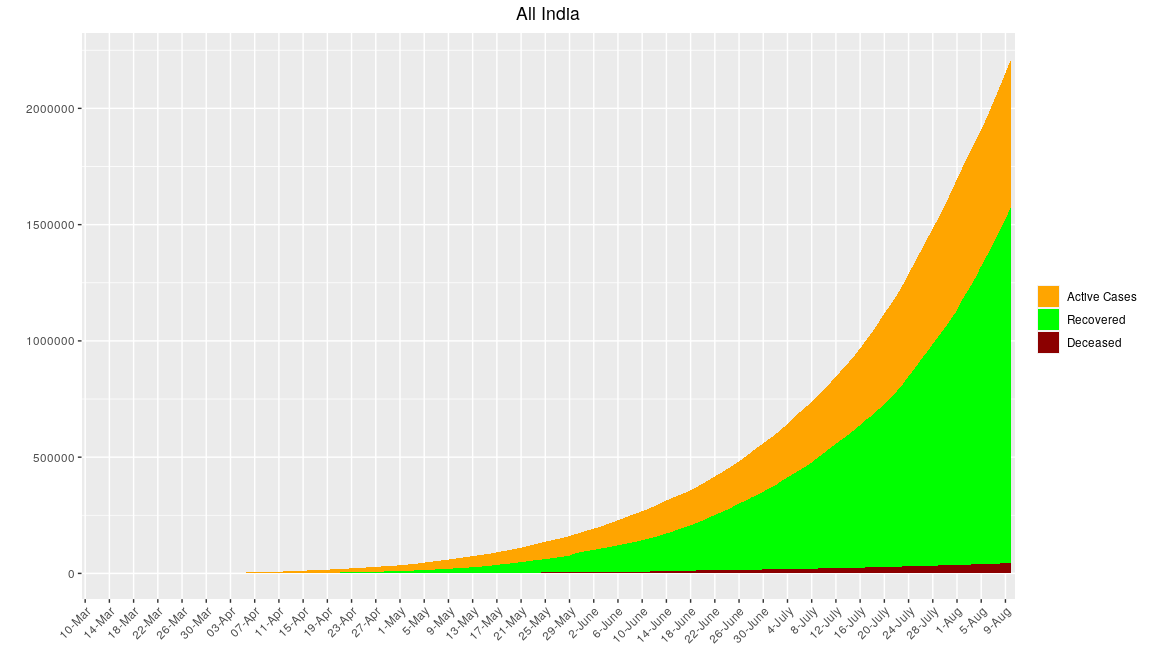}
\caption{\small Timeline of COVID-19 cases, recoveries and fatalities in India taken from~\cite{Siva-website}. See ~\cite{Siva-website} and~\cite{covid19India} for detailed information on how COVID-19 progressed in India.}
\label{fig:India-timeline}
\end{figure*}

COVID-19 is an ongoing pandemic that began in December 2019. The first case in India was reported on 30\,January\,2020. The number of cases and fatalities have been on the rise since then. As on 11 August 2020, there are 22,68,675 cases (of which 15,83,489 have recovered) and 45,257 fatalities~\cite{MoHFW-website}; see Figure~\ref{fig:India-timeline} for a timeline of COVID-19 cases, recoveries and fatalities in India. While medicines/vaccines for treating the disease remained under development at the time of writing this paper, many countries implemented non-pharmaceutical interventions such as testing, tracing, tracking and isolation, and broader approaches such as quarantining of suspected cases, containment zones, social distancing, lockdown, etc. to control the spread of the disease. For instance, the Government of India imposed a nation-wide lockdown from 25\,March\,2020 to 14\,April\,2020, and subsequently extended it until 31\,May\,2020 to break the chain of transmission and also to mobilise resources (increase healthcare facilities and streamline procedures). To evaluate various such interventions and decide which route to take to manage the pandemic, epidemiologists resort to models that predict the total number of cases and fatalities in both the immediate and the distant futures. The models used should have enough features to enable the evaluation of the impact of various kinds of non-pharmaceutical interventions.

\begin{figure*}[t]
\centering
\includegraphics[scale=0.125]{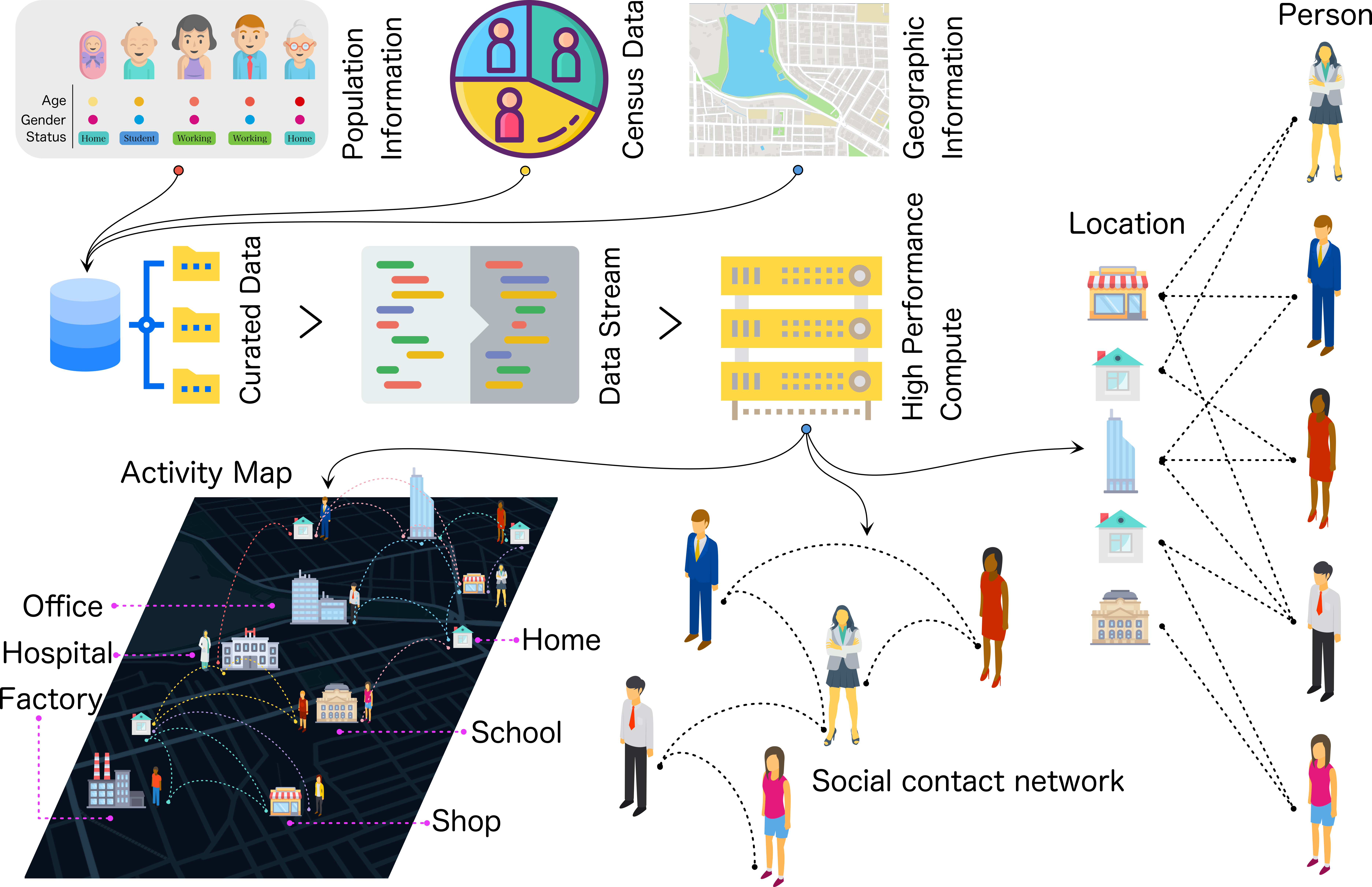}
\caption{\small Schematic representation of an agent-based model.}
\label{fig:abm-art}
\end{figure*}

Broadly three kinds of models have been used to study this epidemic. The first set of models takes a curve-fitting approach. They rely on simple parametric function classes. The parameters of the model are fit via regression to match observed trends. The second set of models addresses the physical dynamics of the spread at a macroscopic level. These are mean-field ordinary differential equations (ODEs) based compartmental models (e.g. Susceptible-Exposed-Infected-Recovered (SEIR) model and its extensions) based on the classical work of Kermack and McKendrick \cite{kermack1927contribution}. Here the population is divided into various compartments such as susceptible, exposed, infected, recovered, etc., based on the characteristics of the epidemic. One then solves a system of ODEs that captures the evolution of the epidemic at a macroscopic scale\footnote{See \cite{INDSCISIM-2020} for a state-level epidemiological model for India and \cite{prakash2020minimal} for a combination of the two approaches.}. Localised versions of these are spatio-temporal mean-field models that lead to partial differential equations\footnote{For a paper in the Indian context see \cite{ganesan2020spatio}.}. The third set of models, and the focus of this work, are agent-based models\footnote{There are other agent-based simulators that have informed policy decisions. See \cite{ferguson2020report} for UK and USA related studies specific to COVID-19,
see \cite{gardner2020intervention} for a COVID-19 study on Sweden, see \cite{halloran2008modeling} and references therein for many agent-based  models and their comparisons, and see \cite{hunter2017taxonomy} for a taxonomy of agent-based models.}. A very detailed model of the society under consideration, with as many agents as the population, is constructed using census and other data. The agents interact in various interaction spaces such as households, schools, workplaces, marketplaces, transport spaces, etc. See Figure~\ref{fig:abm-art} for a schematic representation of an agent-based model with the aforementioned interaction spaces. These interaction spaces are the primary contexts for the spread of infection. A susceptible individual can potentially get infected from an interaction in one of these spaces upon contact with an infected individual. Once an individual is exposed to the virus, this person goes through various stages of the disease, may infect others, and eventually, either recovers or dies. Other models work at an intermediate level by modelling the social network of interactions, e.g., \cite{eubank2004modelling}, but we shall focus more on agent based models.

There are several advantages of using agent-based models. First, since modelling is performed at a microscopic level unlike the macroscopic level in compartmental models, agent-based models are well suited to capture heterogeneity at various levels. For instance, the age-dependent progression of COVID-19 in individuals (severity, the need for hospital care, intensive care, etc.) can be incorporated in agent-based models. Second, individual behavioural changes, known to be important in certain diseases such as AIDS, can be easily modelled. Third, agent-based models are well suited to study the impact of various non-pharmaceutical interventions, such as ``lockdown for a certain number of days", ``offices operating using the (so-called) odd-even strategy", ``social distancing of the elderly", ``voluntary home quarantine", ``closure of schools and colleges", etc. Explicit modelling of these contexts of infection spread also enables studies of control measures targeting the interaction spaces. Fourth, there is an important difference between the \emph{actual infected number} in the population, which is what the differential equations-based models predict, and the \emph{reported cases}. The latter is invariably based on those that come to hospitals/clinics seeking health care, or those that are identified due to random testing, followed by contact tracing of such ‘index’ cases. As a consequence, reported cases provide a \emph{biased} estimate of the actual infected number in the population. Agent-based simulators have the capability to track such biased estimates of prevalence.

In this work, we describe our city-scale agent-based simulator to study the epidemic spread in two Indian cities and demonstrate how digital computational capabilities can help us assess the impact of various interventions and manage a pandemic.

We now provide sample outcomes for Bengaluru and Mumbai for COVID-19 under various interventions. These outcomes have been generated using our city-scale agent-based simulator. Bengaluru and Mumbai have estimated populations of 1.23 and 1.24 crore people respectively\footnote{The 2011 census figure for Bengaluru is 0.85 crore and for Mumbai is 1.24 crore (Mumbai city only, not the Mumbai Urban Area whose 2011 census estimate is 1.84 crore). Bengaluru's 2020 population is estimated to be 1.23 crore. Reliable data is not available for Mumbai city's 2020 population. We have used 2020 estimated population for Bengaluru and 2011 census estimate for Mumbai.}, and our simulator has instantiated those many agents. The Bengaluru population is spread over 198 administrative units called wards. Similarly the Mumbai population is spread over similar administrative units or wards, but there are 24 such wards in Mumbai. Since these are larger wards compared to Bengaluru's wards, there is significant variation of population density within each ward in Mumbai. To model higher spread in densely populated areas, each of the 24 wards is modelled to have subareas with denser population.

\subsubsection{Bengaluru}

\begin{figure*}[t]
\centering
\includegraphics[scale=.46]{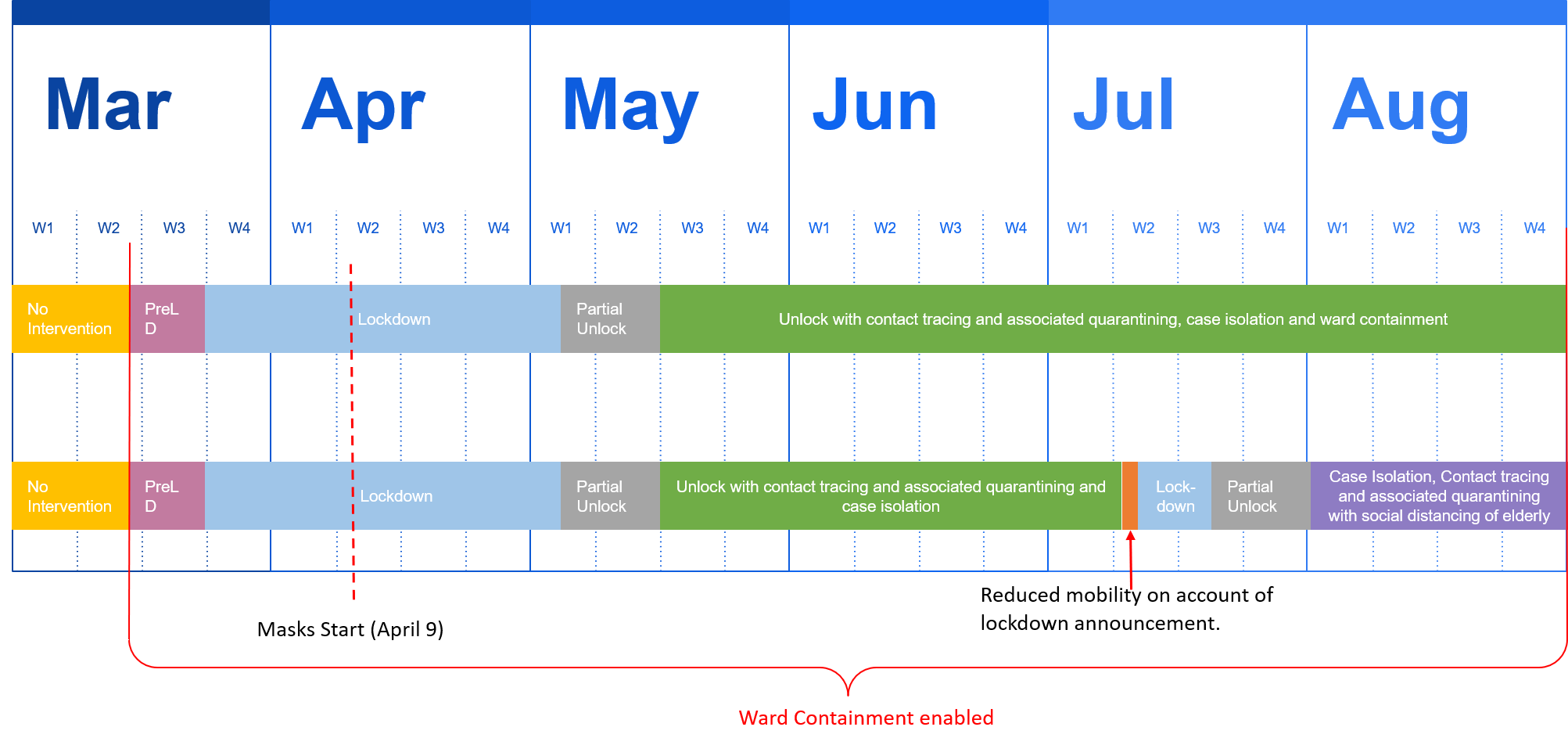}
\caption{\small A timeline of Bengaluru interventions.}
\label{fig:Bengaluru-interventions}
\end{figure*}

\begin{figure}[t]
\centering
\includegraphics[scale=0.5]{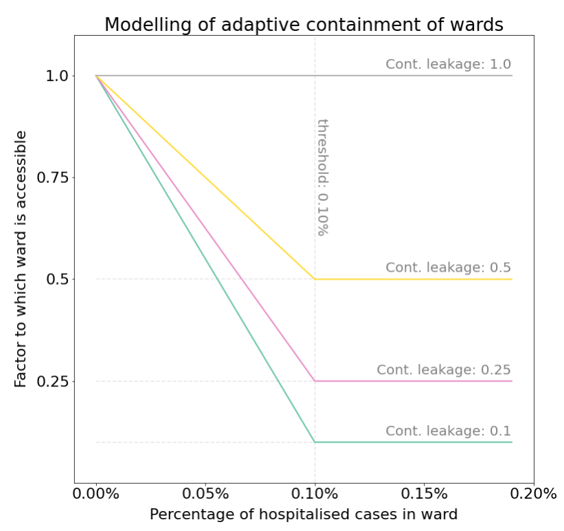}
\caption{\small Wards are contained in a `soft' way. The mobility is gradually decreased based on the signal of number of hospitalised cases in the ward. When 1 in 1000 in the ward is hospitalised, a local lockdown comes into effect.}
\label{fig:soft-containment}
\end{figure}

\begin{figure}[!t]
\centering
\includegraphics[scale=0.4]{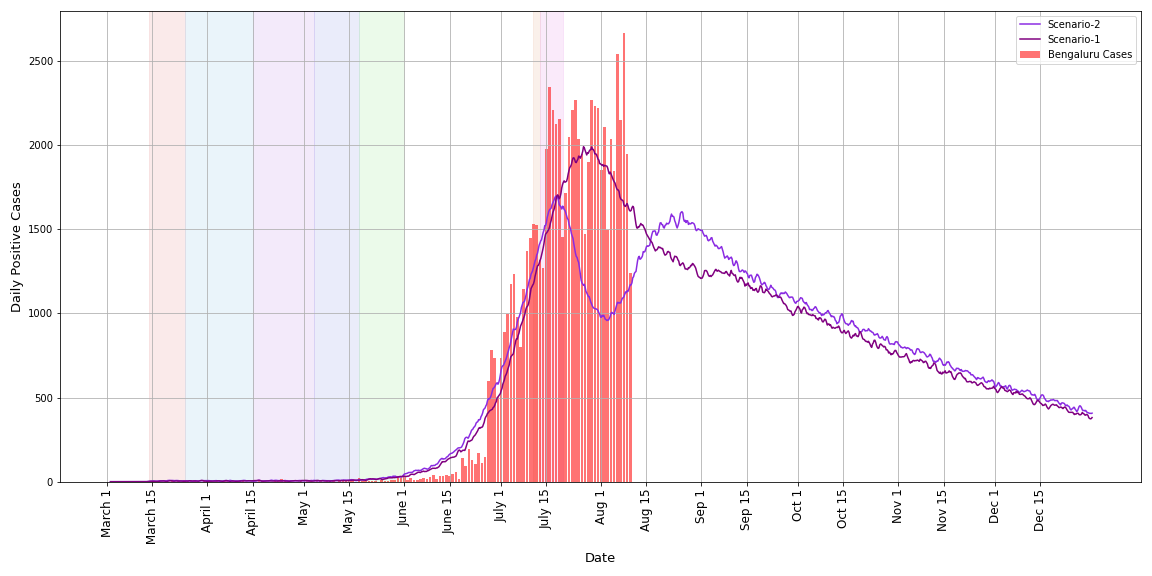}
\caption{\small Bengaluru daily positive cases estimation. The red bars are the reported cases. The five shaded regions between 14\,March and 01\,June represent the durations of the various lockdown phases. The shaded region around 15\,July represents a short one-week lockdown. For cumulative case plots, see Figure \ref{fig:Bengaluru-a-cumulative-cases}.}
\label{fig:Bengaluru-daily-positive-cases-prediction}
\end{figure}

\begin{figure}[t]
\centering
\includegraphics[scale=0.4]{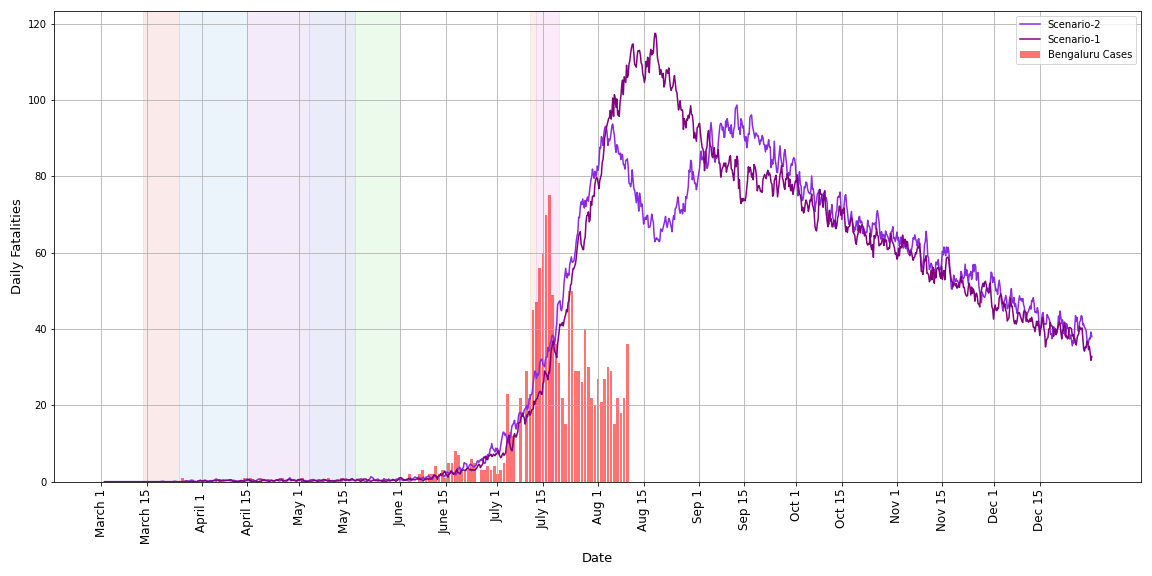}
\caption{\small Bengaluru daily fatalities estimation. For cumulative fatalities, see Figure \ref{fig:Bengaluru-a-cumulative-fatalities} in a later case study.}
\label{fig:Bengaluru-daily-fatalities-prediction}
\end{figure}

For Bengaluru, we consider the following scenario. The Government of India implemented a 40-day lockdown starting 25\,March\,2020. Bengaluru and Karnataka had already closed some interaction spaces in the form of a pre-lockdown. After the 40-day lockdown, there was a phased opening of various activities and offices. More details are as follows. These can also be read from Figure~\ref{fig:Bengaluru-interventions}. We then provide a simulation outcome for these interventions and compare them with the actual situation on the ground. The details:
\begin{itemize}
  \item Pre-lockdown from 14--24\,March\,2020, the first shaded region in Figures~\ref{fig:Bengaluru-daily-positive-cases-prediction}--\ref{fig:Bengaluru-daily-fatalities-prediction}.
  \item 40 days of lockdown from 25\,March -- 03\,May\,2020, the second and the third shaded regions in the figures.
  \item 14 days of phased opening from 04--17 May 2020 involving voluntary home quarantine, social distancing of the elderly, closure of schools and colleges, 50\% occupancy at workplaces, and case isolation. This is the fourth shaded region in the plots.
  \item From 18 May 2020 onwards, continued contact tracing (following the Indian Council of Medical Research (ICMR) guidelines as much as possible) and associated quarantining and case isolations, but otherwise an unlocked Bengaluru. \emph{Soft ward containment} continues to be in force. By soft ward containment, see Figure~\ref{fig:soft-containment}, we mean linearly-varying mobility control that turns an open ward into a locked ward when the number of hospitalised cases become 0.1\% of the ward's population; in the latter locked scenario, only 25\% mobility is allowed for essential services.
  \item Past studies \cite{mask-respiratory2009,maskuse-influenza2010,mask-respiratoryvirus2020,chu2020physical} have indicated that masks have been effective in reducing the spread of influenza. Anecdotal evidence seems to suggest that masks are effective for COVID-19. The Ministry of Home Affairs (MHA) order of 15\,April\,2020 \cite[Annexure~1]{20200415-MHAOrder} made the wearing of masks in public places compulsory. This was re-emphasised in the MHA order of 30\,May\,2020 \cite{20200530-MHAOrder}. We assume that masks are mandatory from 09 April 2020 onwards.
  \item It is often the case that when there are several restrictions in place, only a fraction of the population complies with these restrictions. Getting the entire population to comply is often a big challenge and requires significant and persistent messaging (including communication, rewards, punitive measures). We assume a \emph{compliance factor} of 0.7 up to 04 May 2020, which means that 70 percent of the population adheres to the government guidelines like social distancing, wearing masks in public places, etc., and 0.6 thereafter. The reduction could be attributed to behavioural changes due to lockdown fatigue.
  \item A brief lockdown during 14--21 July 2020 was implemented in Bengaluru. We compare two scenarios, one with this lockdown and one without this lockdown.
\end{itemize}

As one can anticipate, simulation of the above scenarios requires a significant level of sophistication in the modelling and implementation. We describe how we do these in the coming sections, but now focus only on the outcomes.

Figure~\ref{fig:Bengaluru-daily-positive-cases-prediction} estimates the daily positive cases and Figure~\ref{fig:Bengaluru-daily-fatalities-prediction} estimates the daily fatalities directly due to COVID-19 in Bengaluru. In both these plots, we compare our estimates with the situation on the ground. The plots are the means of 5 runs each on two versions of synthetic Bengaluru. The jaggedness is due to the stochasticity associated with the limited number of runs. For greater clarity, we have not included the standard error plots.

Figure~\ref{fig:Bengaluru-daily-positive-cases-prediction} and  Figure~\ref{fig:Bengaluru-daily-fatalities-prediction} provide the estimates with and without this one-week lockdown. The trend for the reported cases is roughly captured, but fatalities are over-predicted. This is surprising since the reported cases continued to be high in the third week of July. For a more detailed study of these plots, we refer the reader to Section \ref{subsec:casestudyA}. At this stage, we only observe that the public health benefit of the lockdown is clear from the pictures, reduced peak at the expense of a brief second wave. Armed with these predicted outcomes under the two scenarios, public health officials can now weigh the benefits of the lockdown against its economic consequences.

\FloatBarrier

\subsubsection{Mumbai}
\begin{figure}[!t]
\centering
\includegraphics[scale=0.4]{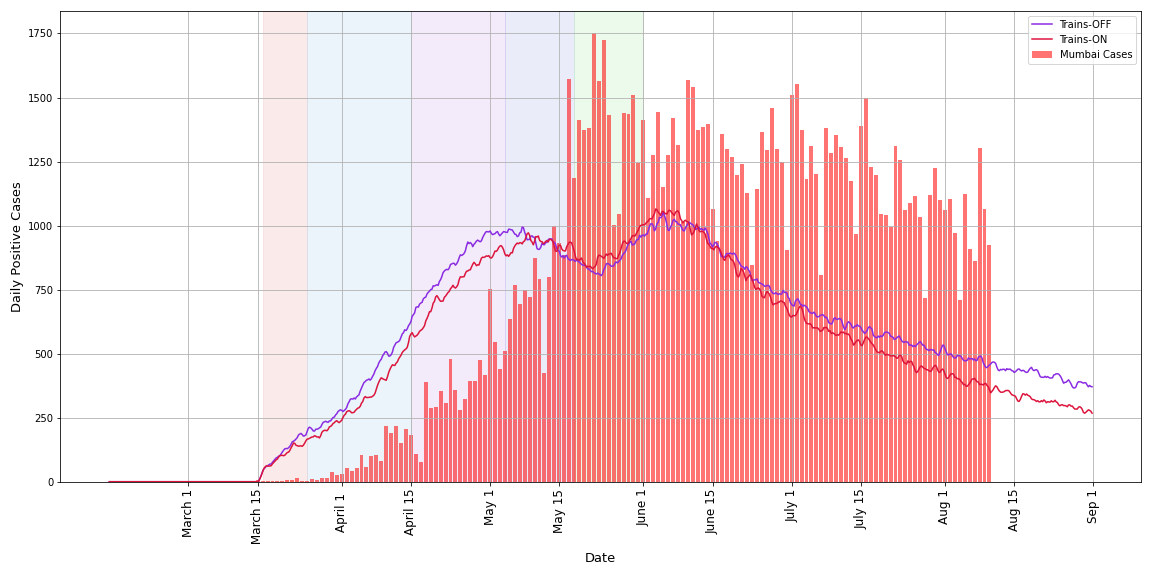}
\caption{\small Mumbai daily positive cases estimation.  The five shaded regions between 16\,March and 01\,June represent the durations of the various lockdown phases in Mumbai. For cumulative case plots, see Figure \ref{fig:Mumbai-c-cumulative-cases}.}
\label{fig:Mumbai-daily-positive-cases-prediction}
\end{figure}

\begin{figure}[t]
\centering
\includegraphics[scale=0.4]{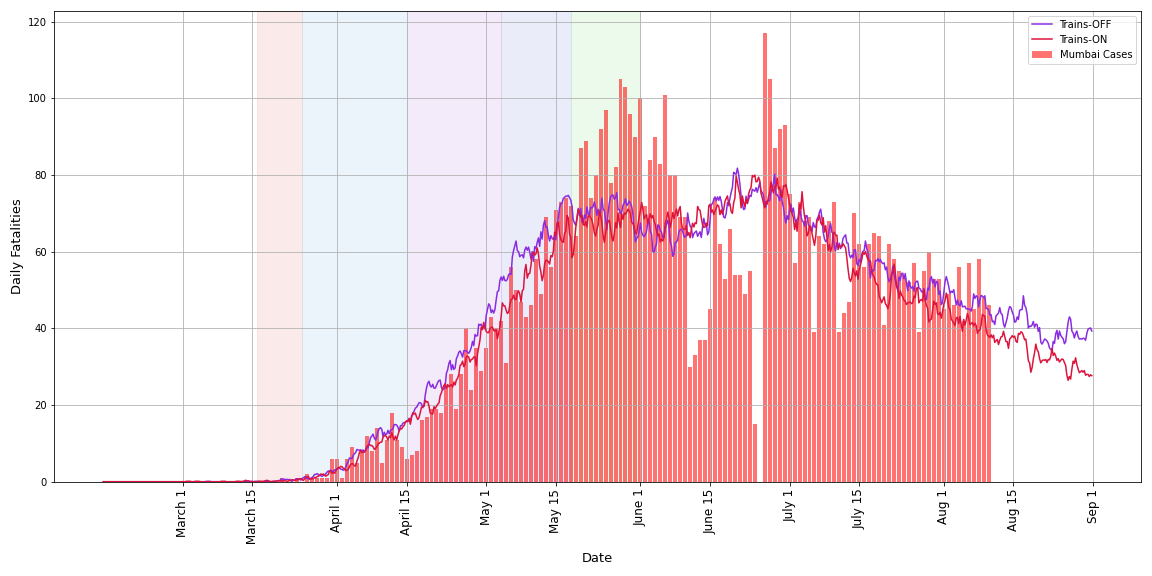}
\caption{\small Mumbai daily fatalities estimation vs. corrected fatality time series, as corrected by BMC on 18 June 2020. For cumulative fatalities, see Figure \ref{fig:Mumbai-c-cumulative-fatalities} in a later case study.}
\label{fig:Mumbai-daily-fatalities-prediction}
\end{figure}

For Mumbai, we simulate the following scenario.
\begin{itemize}
  \item A pre-lockdown similar to Bengaluru, but during 16-24\,March\,2020 (first shaded region in Figure~\ref{fig:Mumbai-daily-positive-cases-prediction}-\ref{fig:Mumbai-daily-fatalities-prediction}).
  \item 40+14 days of lockdown from 25\,March -- 17\,May\,2020, the second, third, and fourth shaded regions in the figures.
  \item Masks are mandatory from 09 April 2020.
  \item Workplaces open with a small strength of 5\% during 18-31 May 2020, as per Government of Maharashtra directions. This is the fifth shaded region. During this period, social distancing of the elderly and school and college closures remain in force.
  \item Workplace strengths increase to 20\% in June, to 33\% in July, and to 50\% in August, with commensurate capacity increases in the local trains. Social distancing of the elderly and school and college closures remain in force. In addition voluntary home quarantine and case isolation come into play.
  \item Throughout the simulations, soft ward containment is in force.
  \item It is often difficult to comply with social distancing directives in high population density areas, like in slums, with many common essential facilities. In Mumbai, we model compliance to be 0.4 in high density areas and 0.6 in other areas.
  \item Throughout the simulations, contact tracing, associated quarantining, testing, and further tracing are enabled.
  \item We will compare the above scenario, with local trains enabled, and will contrast it with another hypothetical scenario having no local trains.
\end{itemize}

Figure~\ref{fig:Mumbai-daily-positive-cases-prediction} estimates the daily positive cases and Figure~\ref{fig:Mumbai-daily-fatalities-prediction} estimates the daily fatalities. In both these plots, we compare our estimates with the situation on the ground. Again the plots are the means of five runs each on two synthetic versions of Mumbai.

But for a delay in the estimated cases curve, the trends for cases and fatalities are captured well. The delay in the estimated cases is perhaps due to delayed reporting which is not modelled in the simulator. From these figures, one can recognise the usefulness of the agent-based simulator in assessing the impact of opening of the local trains in Mumbai.

\vspace*{.5cm}

\noindent \emph{Agent-based simulator -- an important intervention-planning tool}: Let us summarise the outcomes of the above two examples. We saw the  public health impact of imposing a short lockdown in Bengaluru. We also saw the impact of opening up trains vs. keeping the trains non-operational in Mumbai. Such comparisons that can inform better on-ground decision making are enabled by a city-scale agent-based simulator. This capability arises mainly because of the enhanced modelling and control of the interaction spaces in the simulator. It is our hope that such tools become common place in a city administration's tool kit, and are used to the fullest extent before drastic interventions with wide-scale impact, e.g., lockdown, are imposed. With additional modelling of activity, mobility, and behaviour, and use of high quality data on the migrant labour force in urban areas, we speculate that we could have anticipated certain behavioural outcomes seen in India after the lockdown announcement (e.g., migrant population movement).

\section{Methodology: Agent-based modelling}
Broadly, the steps involved in agent-based modelling are the following: build the simulator, calibrate it, validate it, and use it for estimating how the pandemic will evolve.

1. {\bf Simulator}. The simulator itself consists of four parts.

\emph{Synthetic city}. A synthetic city generator builds a synthetic city with individuals and various interaction spaces. Individuals are assigned to various interaction spaces such as households, schools/workplaces, communities and transport spaces. In doing this we capture the demographics of the city, the school size distributions, the workplace size distributions, the commute distances, the neighbourhood and friends' interaction networks, the transport interaction spaces, etc. These fix the ``social networks'' on which individuals interact and transmit the virus.

\emph{Disease progression}. A disease progression model that involves the biology of the disease then indicates what is the incubation period, infective period, symptomatic period, severity of the symptoms, viral load, virus shedding, health care and in-hospital progression, etc.

\emph{Interactions and evolution}. The level of infectivity during the infective period, the duration of the infective period, and the social network interactions in the various interaction spaces determine how the disease evolves in the city. We start the simulator with a certain number of infected individuals. They then interact with susceptible individuals at various interaction spaces, who in turn interact with other susceptible individuals, and thus the epidemic progresses. The key parameters in the disease evolution are the transmission coefficients associated with each interaction space that model the chance of meetings and disease spread in that interaction space.

\emph{Intervention model}. Various kinds of intervention policies need to be defined and their impact on transmission coefficients should be modelled. See Table~\ref{tab:interventions0} for some examples. Many of these involve reduction in changes in contact rates as a consequence of the interventions. The values to set could be based on observed mobility patterns. For example, according to the COVID-19 Community Mobility Report for India in April \cite{20200411mobilitychanges} in Table \ref{tab:mobility reduction}, prepared by Google based on data from Google Account users who have ``opted-in'' to location history, there was significant reduction in mobility during the lockdown period compared to the baseline period of 03\,January\,2020 to 06\,February\,2020. This informs the nominal contact rate choices in the interventions' definitions in Table \ref{tab:interventions0} and later in other Tables.

\begin{table*}[!ht]
\caption{Intervention modelling}
\label{tab:interventions0}
\begin{center}
\begin{tabular}{||p{0.1\textwidth}|p{0.23\textwidth}|p{0.6\textwidth}||}
 \hline
 \hline
 Label & Policy & Description \\
 [0.5ex]
 \hline
 \hline
 CI & Case isolation at home & Symptomatic individuals stay at home for 7 days, non-household contacts reduced by 90\% during this period, household contacts reduce by 25\%.\\
 \hline
 HQ & Voluntary home quarantine & Once a symptomatic individual has been identified, all members of the household remain at home for 14 days. Non-household contacts reduced by 90\% during this period, household contacts reduce by 25\%. \\
 \hline
 SDO & Social distancing of those aged 65 and over & Non-household contacts reduce by 75\%. \\
 \hline
 LD & Lockdown & Closure of schools and colleges. Only essential workplaces active. For a compliant household, household contact rate doubles, community contact rate reduces by 75\%, workspace contact rate reduces by 75\%. For a non-compliant household, household contact rate increases by 25\%, workspace contact rate reduces by 75\%, and no change to community contact rate.\\
 \hline
 \hline
\end{tabular}
\end{center}
\end{table*}

\begin{table}[!ht]
\caption{Mobility report generated on 11 April 2020, see \cite{20200411mobilitychanges}.}
\label{tab:mobility reduction}
\begin{center}
\begin{tabular}{||p{0.3\textwidth}|p{.1\textwidth}||}
 \hline
 \hline
 Place & Reduction \\
 [0.5ex]
 \hline
 \hline
 Retail and recreation &  -80\% \\
 Grocery and pharmacy & -55\% \\
 Parks and public plazas & -52\% \\
 Public transit stations & -69\% \\
 Workplaces & -64\% \\
 Residential & +30\% \\
 \hline
 \hline
\end{tabular}
\end{center}
\end{table}

2. {\bf Calibration}. Once the simulator is ready there are still unknown parameters that need to be identified. These include the contact rates at various interaction spaces, the number of infections to seed, the time at which these infections should be seeded, the compliance parameters, etc. The purpose of the calibration step is to identify these parameters to capture the city specific trends and contact rates. We do this by choosing the initial number to seed, the time at which these are seeded, and the contact rates so that the initial trend of the disease is matched. Once calibrated, we can run our simulator for a certain number of days and understand how the epidemic spreads.

3. {\bf Validation}. We next have to validate our simulator, so that we can understand the predictive power of the simulator. For this, we look for phenomena in the real data that have not been explicitly modelled and we check if the simulator is able to capture these phenomena. For specific details, see Section~\ref{sec:Simulator}.

4. {\bf Use of the simulator in an evolving pandemic}. It is often the case that in evolving pandemics, predictions do not match reality as time unfolds. Models are often gross over-simplifications of the underlying complex reality and assumptions are often wrong or may need updating as the pandemic evolves. The purpose of models in an evolving pandemic is not merely to predict numbers, in which task they will likely fail, but more to enable principled decision making on intervention strategies. They enable a study of the public health outcomes of one strategy versus another. Armed with these comparisons, public health officials can make more informed decisions. Needless to say, these are often more complex and involve several aspects beyond just public health, e.g. economy, psychology, education, political climate, to name a few\footnote{For a proposal on how to simulate economic and public-health aspects together, see \cite{chandak_epidemiologically_2020}.}.

\FloatBarrier

\section{Design of interventions via case studies}

One of the powerful features of the agent-based simulator is its ability to explicitly control various interaction spaces and study the outcomes. We demonstrate this feature via the case studies for Bengaluru and Mumbai listed in Table~\ref{tab:case-studies}.
\begin{table}[ht]
\caption{Case studies for Bengaluru and Mumbai}
\label{tab:case-studies}
\centering
\begin{tabular}{||p{0.16\textwidth}|p{0.52\textwidth}|p{0.16\textwidth}||}
 \hline
 Case Study & Title     & Section \\
\hline
\hline
Case Study A &  No intervention (but only contact tracing-based isolation) versus lockdown versus the current scenario in Bengaluru & Section~\ref{subsec:casestudyA} \\
\hline
Case Study B & Impact of opening offices at 50\% capacity with higher compliance versus lockdown at lower compliance, a Bengaluru study & Section~\ref{subsec:casestudyB} \\
\hline
Case Study C & Impact of opening trains versus not opening trains in Mumbai & Section~\ref{subsec:casestudyC} \\
\hline
Case Study D & Soft ward containment versus neighbourhood containment in Bengaluru & Section~\ref{subsec:casestudyD} \\
\hline
Case Study E & Soft ward containment at various levels in Mumbai & Section~\ref{subsec:casestudyE} \\
\hline
Case Study F & Schools/colleges open from 01 September 2020 in Bengaluru & Section~\ref{subsec:casestudyF} \\
\hline
\hline
\end{tabular}
\end{table}

\subsection{Case Study A: No intervention (but only contact tracing-based isolation) versus lockdown versus the current scenario in Bengaluru}
\label{subsec:casestudyA}
We compare the following three scenarios in Bengaluru:
\begin{itemize}
\item No intervention other than contact tracing, testing and associated case isolation.
\item Indefinite lockdown starting from 14\,March\,2020 onwards. This naturally will have enormous economic and societal cost, but we focus only on the direct COVID-19 public health outcomes.
\item Scenario-2 in Table~\ref{tab:Bengaluru-interventions}: soft ward containment, case isolation with testing and contact tracing, and a one-week lockdown during 14--21 July 2020.
\end{itemize}

\begin{table}[ht]
\caption{Simulated Bengaluru interventions}
\label{tab:Bengaluru-interventions}
\centering
\begin{tabular}{||p{0.22\textwidth}|p{0.3\textwidth}|p{0.3\textwidth}|p{0.1\textwidth}||}
 \hline
 Period & Scenario-1       & Scenario-2 &  Compliance \\
 \hline
 \hline
 01 -- 13\,March\,2020  & No intervention & No intervention & NA  \\
 \hline
 14 -- 24\,March\,2020 & Prelockdown  & Prelockdown & 70\% \\
 \hline
 25\,March -- 03\,May\,2020 & 40 days of National lockdown & 40 days of National lockdown & 70\% \\
 \hline
 09 April 2020 -- onwards & Masks ON & Masks ON & 60\% \\
 \hline
 04 -- 17 May 2020 & Phased opening.
Voluntary home quarantine, social distancing of elderly, case isolation, schools and colleges closed, 50\% occupancy at workplaces. & Phased opening.
Voluntary home quarantine, social distancing of elderly, case isolation, schools and colleges closed, 50\% occupancy at workplaces.  & 60\% \\
  \hline
 18 May -- 11 July 2020 & Unlocked Bengaluru with only ICMR-guideline contact tracing and associated quarantining and case isolations, social distancing of elderly, case isolation, schools and colleges closed. & Unlocked Bengaluru with only ICMR-guideline contact tracing and associated quarantining and case isolations, social distancing of elderly, case isolation, schools and colleges closed. & 60\% \\
  \hline
 12 -- 13 July 2020 & Same as above & Prelockdown - 50\% mobility & 60\% \\
  \hline
 14 July --  21 July  2020 & Same as above & Bengaluru lockdown & 60\% \\
  \hline
 22 July -- 31 July  2020  & Same as above & Case isolation, social distancing of elderly, school closure, workplaces at 50\% & 60\% \\
  \hline
 01 August 2020 -- onwards & Same as above & Case isolation, social distancing of elderly, school closure, workplaces at 100\% & 60\% \\
 \hline
 Throughout & Soft ward containment enabled & Soft ward containment enabled & 60\% \\
 \hline
\end{tabular}
\end{table}

\begin{figure}
\centering
\includegraphics[scale=0.4]{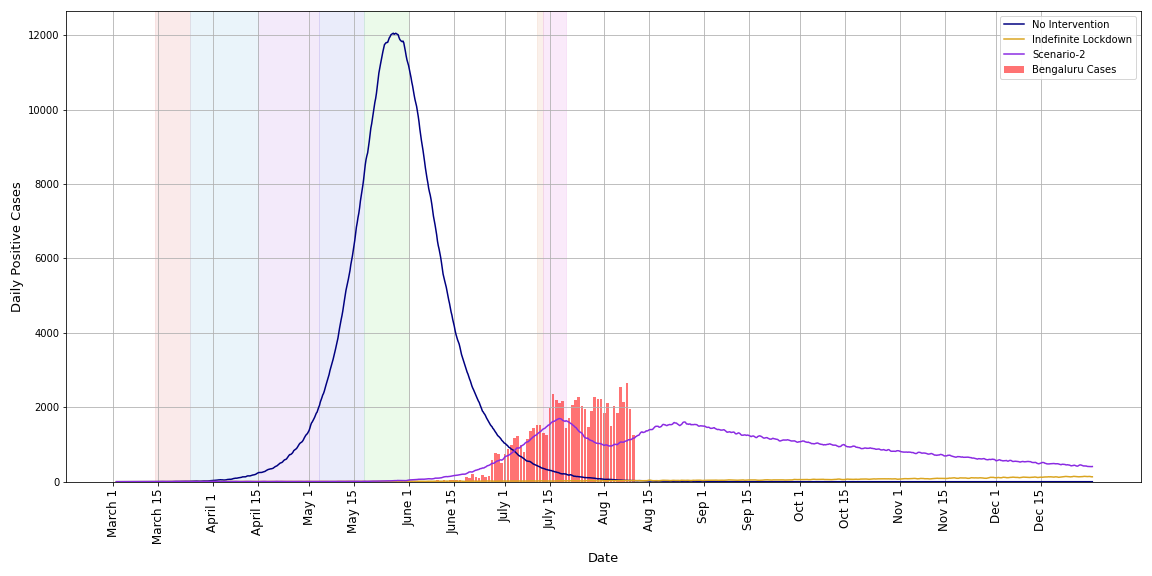}
\caption{\small Case study A, subsection~\ref{subsec:casestudyA}: Bengaluru daily cases estimation. For a magnified view of the lower part of the plot, see Figure~\ref{fig:Bengaluru-daily-positive-cases-prediction}. The no-intervention situation would have overwhelmed the healthcare system many times over.}
\label{fig:Bengaluru-a-daily-cases}
\end{figure}

\begin{figure}
\centering
\includegraphics[scale=0.4]{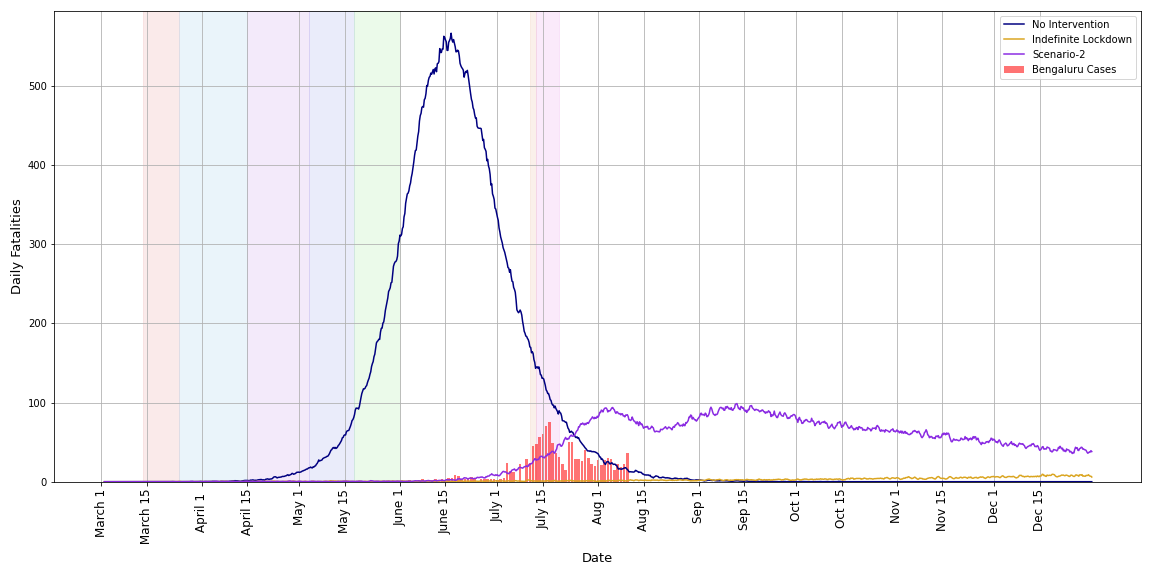}
\caption{\small Case study A, subsection~\ref{subsec:casestudyA}: Bengaluru daily fatalities estimation. For a magnified view of the lower part of the part, see Figure~\ref{fig:Bengaluru-daily-fatalities-prediction}.}
\label{fig:Bengaluru-a-daily-fatalities}
\end{figure}

\begin{figure}
\centering
\includegraphics[scale=0.4]{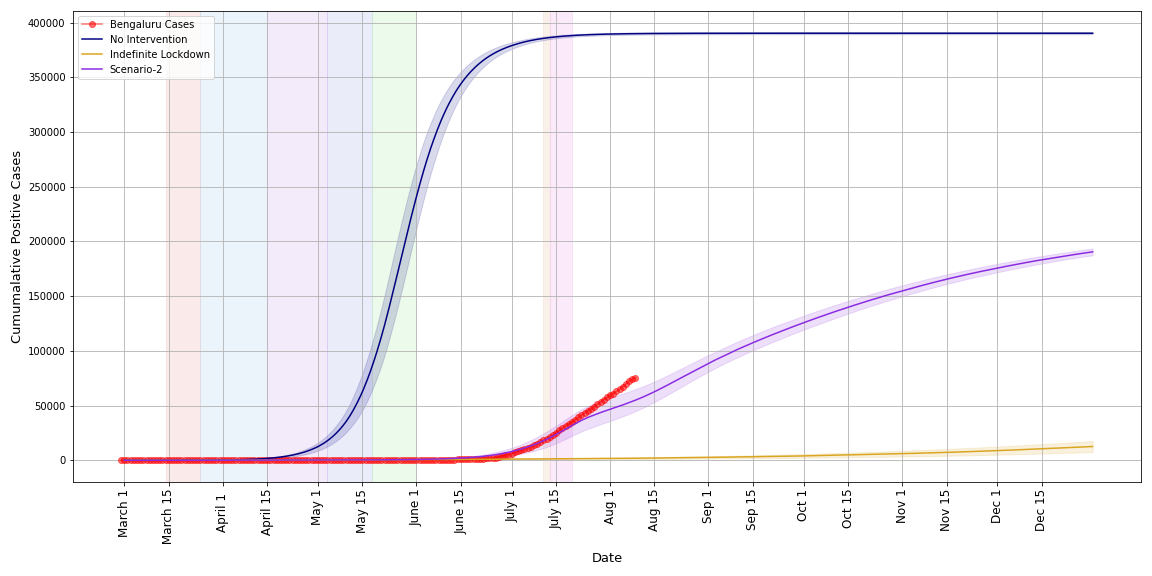}
\caption{\small Case study A, subsection~\ref{subsec:casestudyA}: Bengaluru cumulative cases estimation.}
\label{fig:Bengaluru-a-cumulative-cases}
\end{figure}

\begin{figure}
\centering
\includegraphics[scale=0.4]{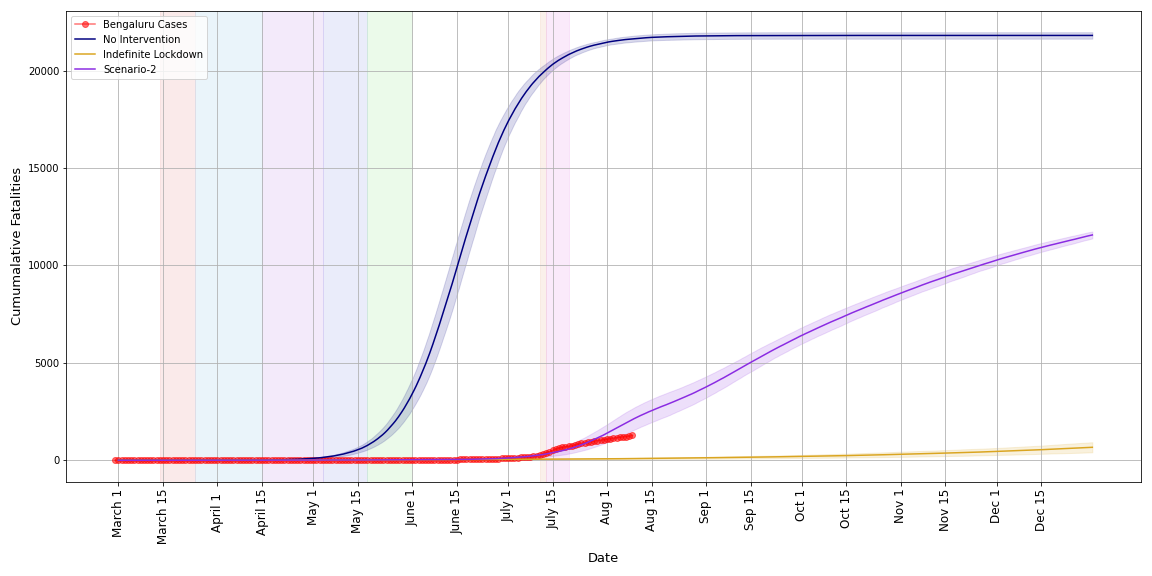}
\caption{\small Case study A, subsection~\ref{subsec:casestudyA}: Bengaluru cumulative fatalities estimation.}
\label{fig:Bengaluru-a-cumulative-fatalities}
\end{figure}

\begin{figure}
\centering
\includegraphics[scale=0.4]{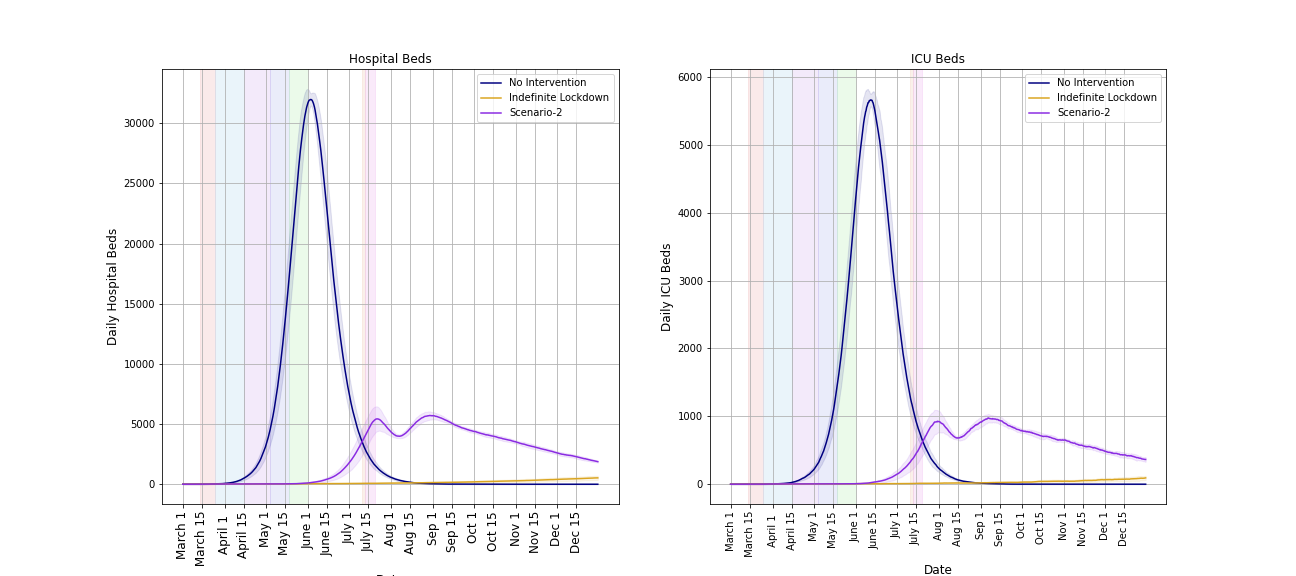}
\caption{\small Case study A, subsection~\ref{subsec:casestudyA}: Bengaluru hospital beds estimation. `Hospital Beds' refers to the number of beds occupied for regular care including possibly oxygen support. `ICU Beds' refers to those that need intensive care or ventilation. The no-intervention scenario would have overwhelmed Bengaluru's healthcare system.}
\label{fig:Bengaluru-a-hospital-beds}
\end{figure}

We assume a compliance of 70\% until 03 May 2020 (i.e. during the initial Karnataka-wide lockdown followed by the nation-wide lockdown) and a compliance of 60\% starting 04 May 2020, for all these scenarios. That is, 70\% (resp.~60\%) of the population comply with the restrictions in place until 03 May 2020 (resp.~starting 04 May 2020). Under these scenarios, we plot the following: daily cases (Figure~\ref{fig:Bengaluru-a-daily-cases}), daily fatalities (Figure~\ref{fig:Bengaluru-a-daily-fatalities}), cumulative cases (Figure~\ref{fig:Bengaluru-a-cumulative-cases}), cumulative fatalities (Figure~\ref{fig:Bengaluru-a-cumulative-fatalities}) and estimated hospital beds and critical care beds (Figure~\ref{fig:Bengaluru-a-hospital-beds}). We make the following observations.
\begin{itemize}
\item As one would expect, the least number of cases, fatalities and hospital beds requirements correspond to the ``indefinite lockdown" scenario. However this scenario has serious impact on the economy, livelihoods, etc.
\item In terms of the daily number of cases, the no intervention scenario had a peak around 01 June 2020 (with roughly 15,000 cases), whereas the present scenario in Bengaluru (i.e. Scenario-2 in Table~\ref{tab:Bengaluru-interventions}) had a much lower peak around 15 July 2020 (with around 2000 cases), followed by another peak around end of August. Similar trends can be seen in the fatalities estimates as well as the hospital bed estimates. Our health care system would have struggled with the no intervention scenario, and the present scenario in Bengaluru helped mitigate and delay the peak of the epidemic.
\item The second predicted peak in Scenario-2 in Table~\ref{tab:Bengaluru-interventions} is due to the one-week lockdown during 14--21 July 2020.
\item Towards the end of July, we overpredict the number of daily fatalities and underpredict the number of daily cases. This could be because of two reasons:
\begin{enumerate}
\item The number of tests has increased significantly during mid-July due to which there is a likely surge in the number of asymptomatic cases. As a consequence, a reduction in the number of daily cases due to the one-week lockdown during 14--21 July is not observed in the reported number of daily cases; such a reduction is visible in our estimates because the testing regime is assumed constant through the period in our simulator.
\item There is a delay in reporting the fatalities. As the reported number of daily cases follow an exponential trend during early-mid July, one would expect a similar trend in the reported daily fatalities during end-July, as shown in our prediction of the daily fatalities under Scenario-2. However, we see a reduction in the reported number of daily fatalities during after 15 July 2020. This could be due to a possible delay in reporting the daily fatalities, or an effective use of the rapid point-of-care antigen test kits, or a combination of both. Testing of these hypotheses require further investigation.
\end{enumerate}
\end{itemize}

\FloatBarrier

\subsection{Case Study B: Impact of opening offices at 50\% capacity with higher compliance versus lockdown at lower compliance}
\label{subsec:casestudyB}

\begin{figure}
\centering
\includegraphics[scale=0.4]{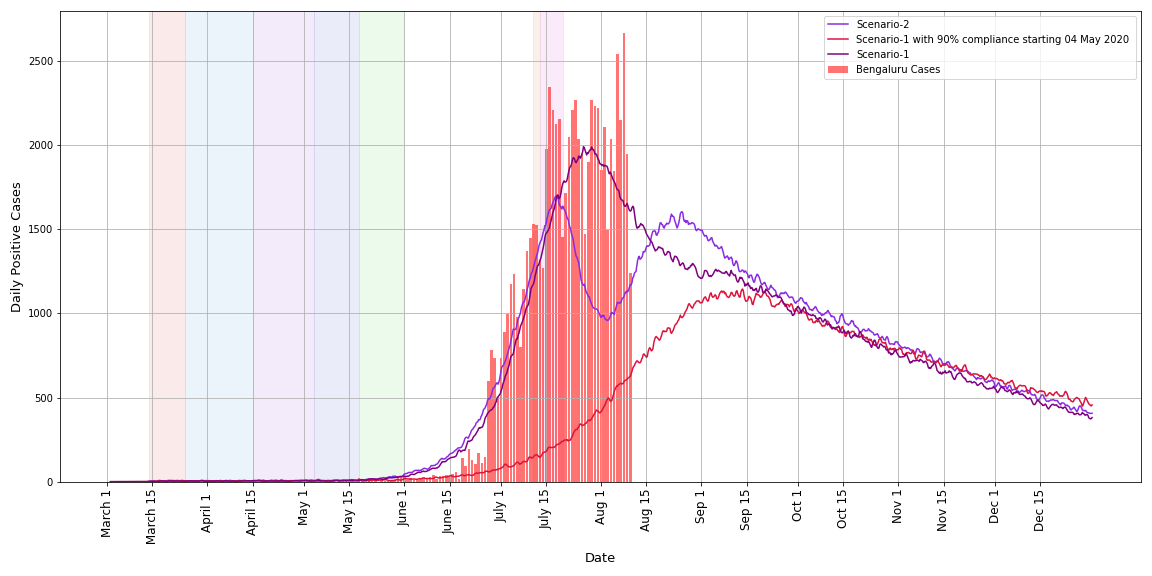}
\caption{\small Case study B, subsection~\ref{subsec:casestudyB}: Bengaluru daily cases estimation.}
\label{fig:Bengaluru-b-daily-cases}
\end{figure}

\begin{figure}
\centering
\includegraphics[scale=0.4]{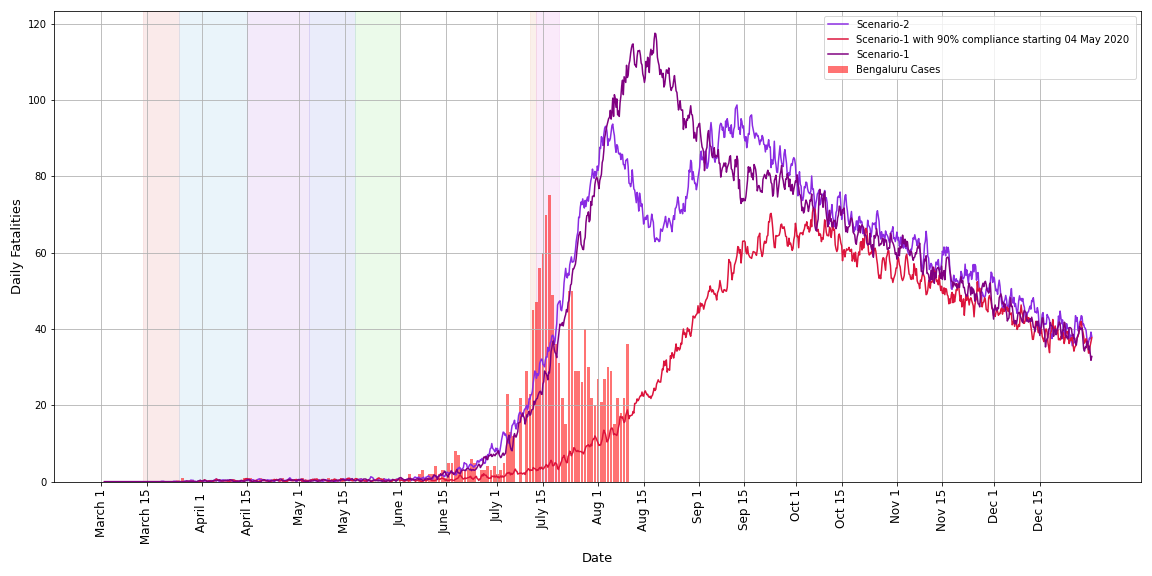}
\caption{\small Case study B, subsection~\ref{subsec:casestudyB}: Bengaluru daily fatalities estimation.}
\label{fig:Bengaluru-b-daily-fatalities}
\end{figure}

\begin{figure}
\centering
\includegraphics[scale=0.4]{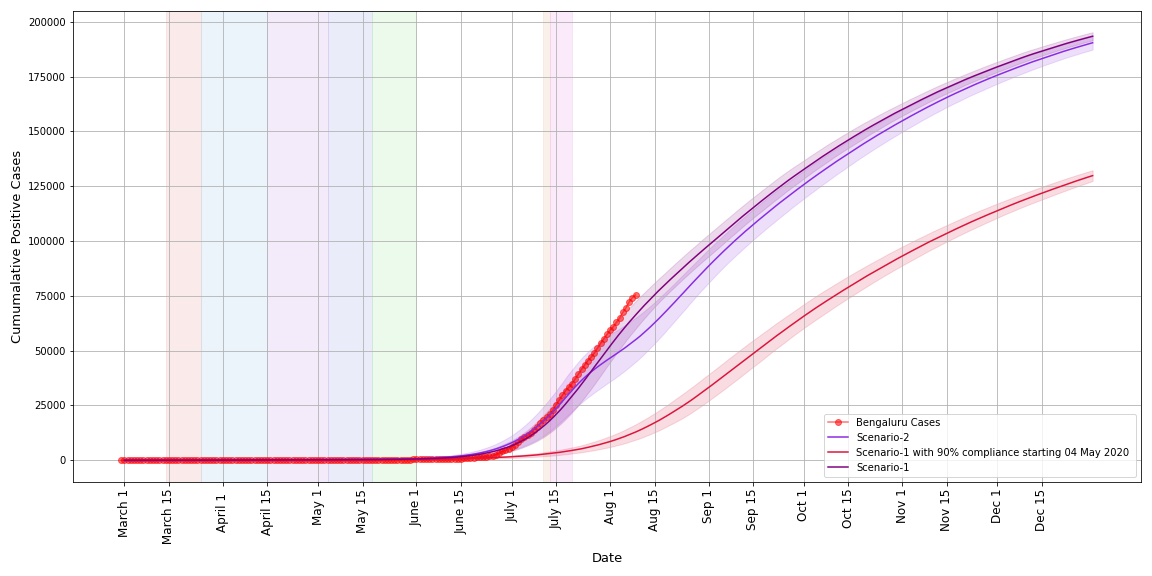}
\caption{\small Case study B, subsection~\ref{subsec:casestudyB}: Bengaluru cumulative cases estimation.}
\label{fig:Bengaluru-b-cumulative-cases}
\end{figure}

\begin{figure}
\centering
\includegraphics[scale=0.4]{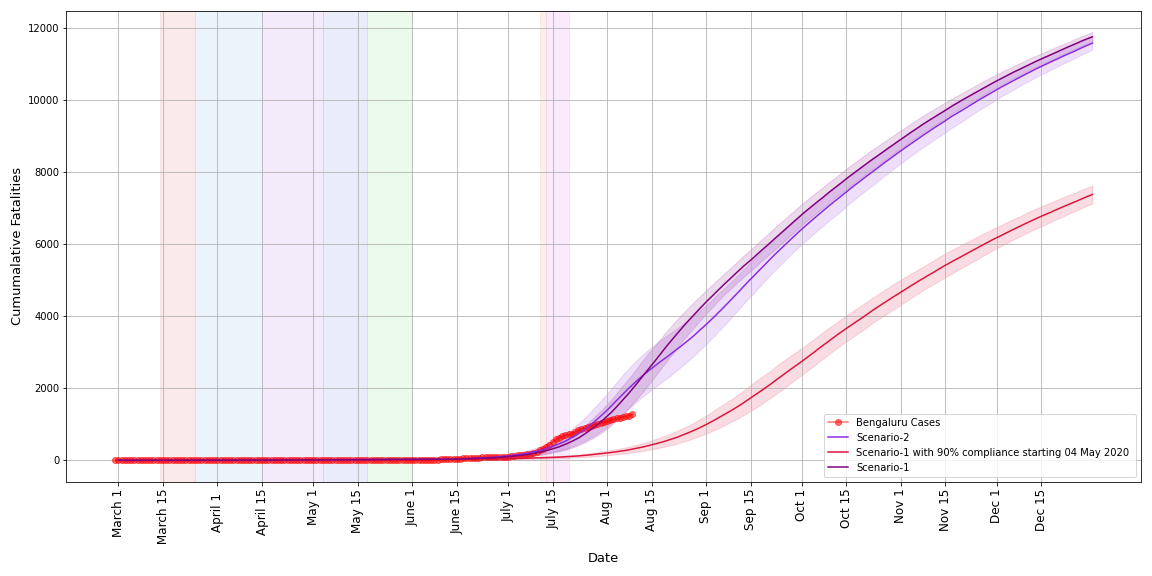}
\caption{\small Case study B, subsection~\ref{subsec:casestudyB}: Bengaluru cumulative fatalities estimation.}
\label{fig:Bengaluru-b-cumulative-fatalities}
\end{figure}

\begin{figure}
\centering
\includegraphics[scale=0.4]{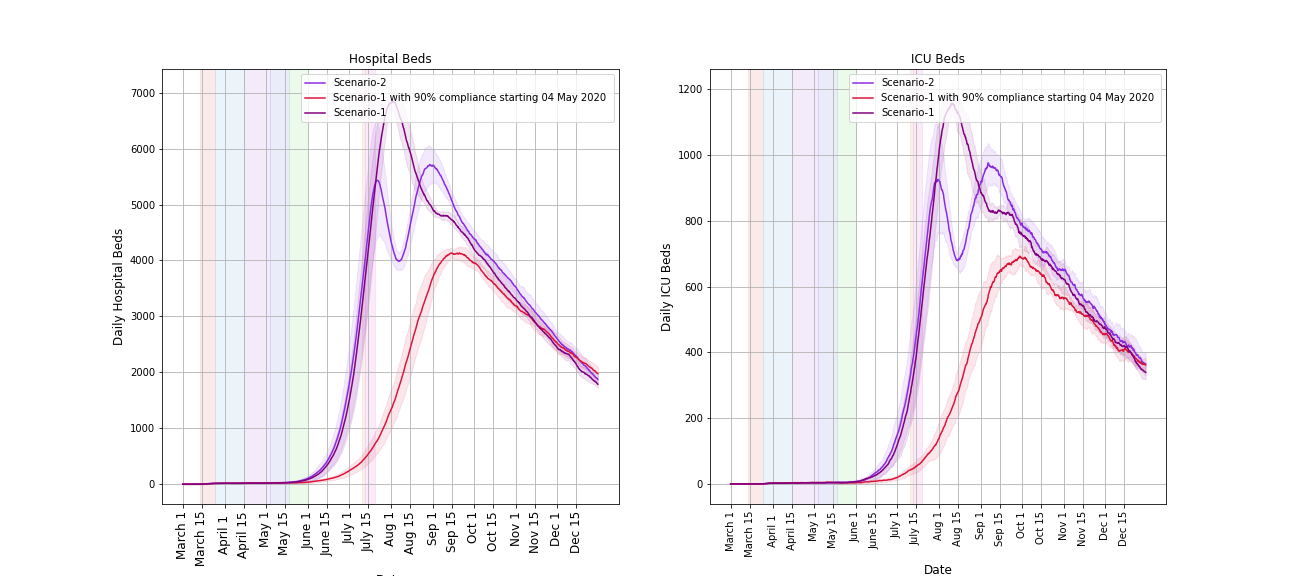}
\caption{\small Case study B, subsection~\ref{subsec:casestudyB}: Bengaluru hospital beds estimation.}
\label{fig:Bengaluru-b-hospital-beds}
\end{figure}

The degree of compliance among the population to public health directions/guidelines is an important factor that affects the epidemic. To understand the importance of compliance, we compare the following scenarios for Bengaluru: the present Bengaluru (i.e. Scenario-2 in Table~\ref{tab:Bengaluru-interventions}), an unlocked Bengaluru (i.e. Scenario-1 in Table~\ref{tab:Bengaluru-interventions}), and an unlocked Bengaluru  with a higher compliance of 90\% starting 04 May 2020 (i.e. Scenario-1 in Table~\ref{tab:Bengaluru-interventions} with 70\% compliance during 14\,March\,2020 -- 03\,May\,2020 and 90\% compliance starting 04 May 2020). As before, we plot the following:  daily cases (Figure~\ref{fig:Bengaluru-b-daily-cases}), daily fatalities (Figure~\ref{fig:Bengaluru-b-daily-fatalities}), cumulative cases (Figure~\ref{fig:Bengaluru-b-cumulative-cases}), cumulative fatalities (Figure~\ref{fig:Bengaluru-b-cumulative-fatalities}) and estimated hospital beds and critical care beds (Figure~\ref{fig:Bengaluru-b-hospital-beds}). We make two important observations:
\begin{itemize}
\item In terms of the number of cases and fatalities, the present Bengaluru (i.e. Scenario-2 in Table~\ref{tab:Bengaluru-interventions}) with 60\% compliance starting 04 May 2020 is worse than an unlocked Bengaluru with 90\% compliance starting 04 May 2020 (with both scenarios having 70\% compliance until 03 May 2020). While the qualitative outcome is not surprising, the quantitative estimates suggest just how important compliance is in curbing the spread of the disease. Armed with such comparisons, city administrations could suitably allocate resources for communication, awareness, and other such campaigns to educate the general populace on the public health impact of their actions, to induce more pro-social behaviour, and to ensure greater compliance. This was the approach taken by Sweden, a country with a population of about 1 crore.
\item Comparing Scenario-1 and Scenario-2, we see that the effect of the one-week lockdown during 14--21 July is very minimal in the long term as far as the cumulative number of cases and fatalities are concerned. However, there is a significant difference in the cumulative number of cases and fatalities between Scenario-2 and Scenario-1 with a higher compliance of 90\% starting 04 May 2020. This suggests that, given that vaccines for COVID-19 are not yet available, short-term lockdowns' benefit is restricted to mobilising resources and preparing the healthcare system in the short term. On the other hand, higher compliance has a greater impact in reducing cases and fatalities.
\end{itemize}

\FloatBarrier

\subsection{Case Study C: Impact of opening trains versus not opening trains in Mumbai}
\label{subsec:casestudyC}

\begin{table}[ht]
\caption{Simulated Mumbai interventions}
\label{tab:Mumbai-interventions}
\centering
\begin{tabular}{||p{0.22\textwidth}|p{0.2\textwidth}|p{0.2\textwidth}|p{0.1\textwidth}|p{0.1\textwidth}||}
 \hline
Period                    & Interventions                                                                                        & Attendance at workplaces& Compliance in non-slums & Compliance in slums \\ \hline
01\,March -- 15\,March\,2020 & No Intervention                                                                                      & at 100\% capacity                & 60\%                & 40\%                 \\ \hline
16\,March -- 08\,April\,2020 & Lockdown                                                                                             &  Essential services operate at 100\% capacity, others are closed                 & 60\%                & 40\%                 \\ \hline
09\,April -- onwards & Masks ON                   &               &                &               \\ \hline
09\,April -- 30\,April\,2020 & Lockdown                                                           & Essential services operate at 100\% capacity, others are closed                 & 60\%           & 40\%              \\ \hline
01\,May -- 17\,May\,2020     & Lockdown with social distancing of the elderly                                                       &  Essential services operate, others are closed                 & 60\%                & 40\%                \\ \hline
18\,May -- 31 \,May\,2020     & Social distancing of the elderly, school closure, community\_factor=0.75                             & Essential services operate at 100\%, others operate at 5\% capacity                  & 60\%                & 40\%                \\ \hline
01 -- 30\,June\,2020   & Home quarantine, social distancing of the elderly, school closure, trains ON, community\_factor=0.75 & Essential services operate at 100\%, others operate at 20\% capacity                     & 60\%                 & 40\%                \\ \hline
01 -- 31\,July\,2020   & Same as above                                                                                        & Essential services operate at 100\%, others operate at 33\% capacity                   & 60\%                 & 40\%                \\ \hline
01\,August  onwards           & Same as above                                                                                        & Essential services operate at 100\%, others operate at 50\% capacity                     & 60\%              & 40\%                \\ \hline
\textbf{Throughout} & \textbf{Soft ward containment enabled} & & & \\
\hline
\end{tabular}
\end{table}

\begin{figure}
\centering
\includegraphics[scale=0.4]{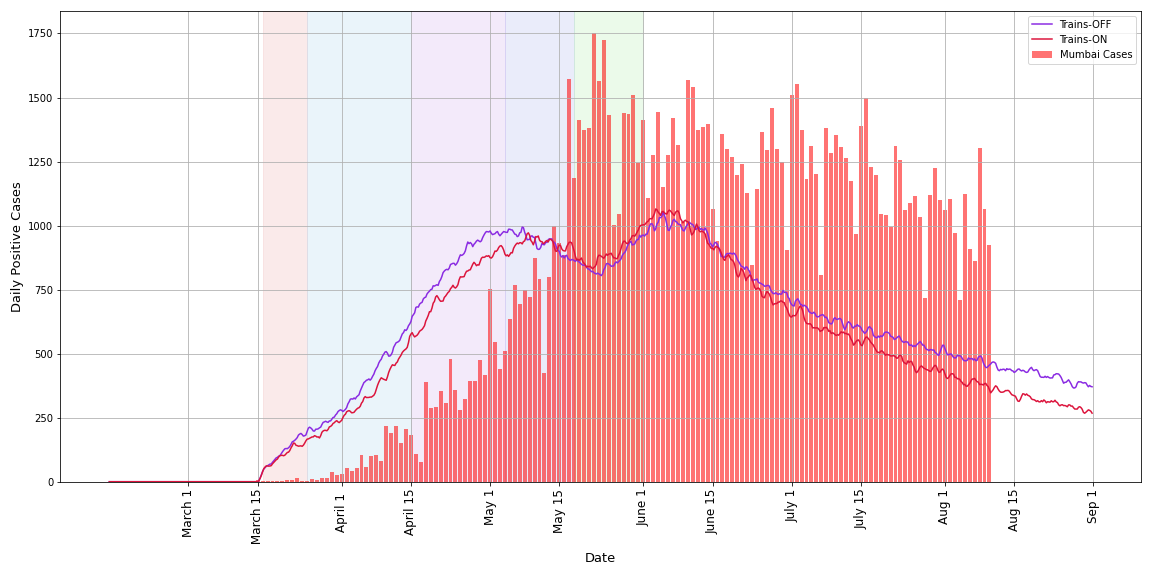}
\caption{\small Case study C, subsection~\ref{subsec:casestudyC}: Mumbai daily cases estimation.}
\label{fig:Mumbai-c-daily-cases}
\end{figure}

\begin{figure}
\centering
\includegraphics[scale=0.4]{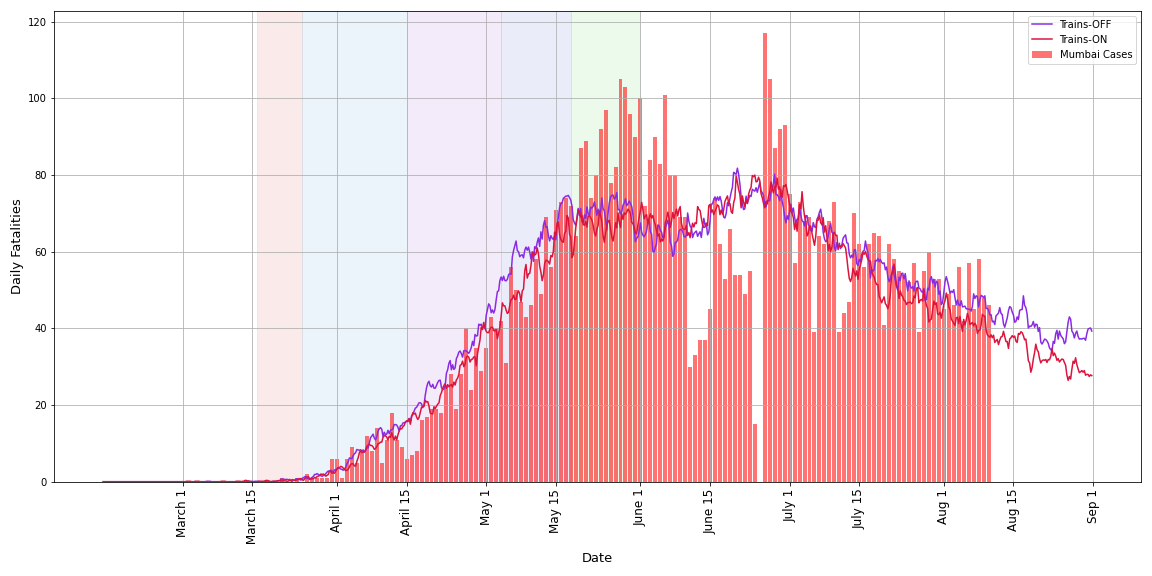}
\caption{\small Case study C, subsection~\ref{subsec:casestudyC}: Mumbai daily fatalities estimation.}
\label{fig:Mumbai-c-daily-fatalities}
\end{figure}

\begin{figure}
\centering
\includegraphics[scale=0.4]{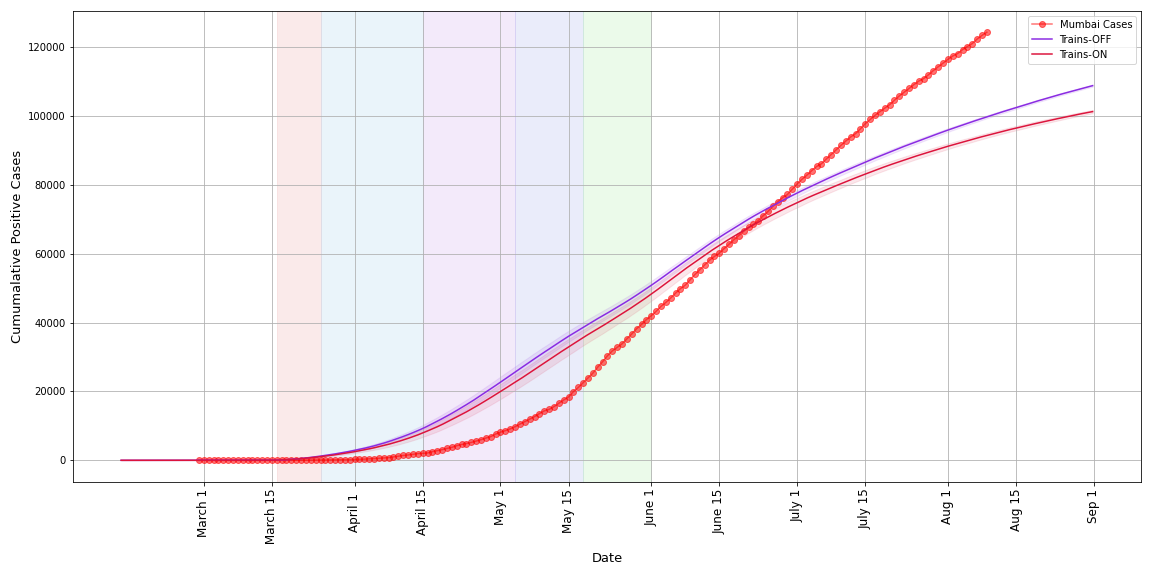}
\caption{\small Case study C, subsection~\ref{subsec:casestudyC}: Mumbai cumulative cases estimation.}
\label{fig:Mumbai-c-cumulative-cases}
\end{figure}

\begin{figure}
\centering
\includegraphics[scale=0.4]{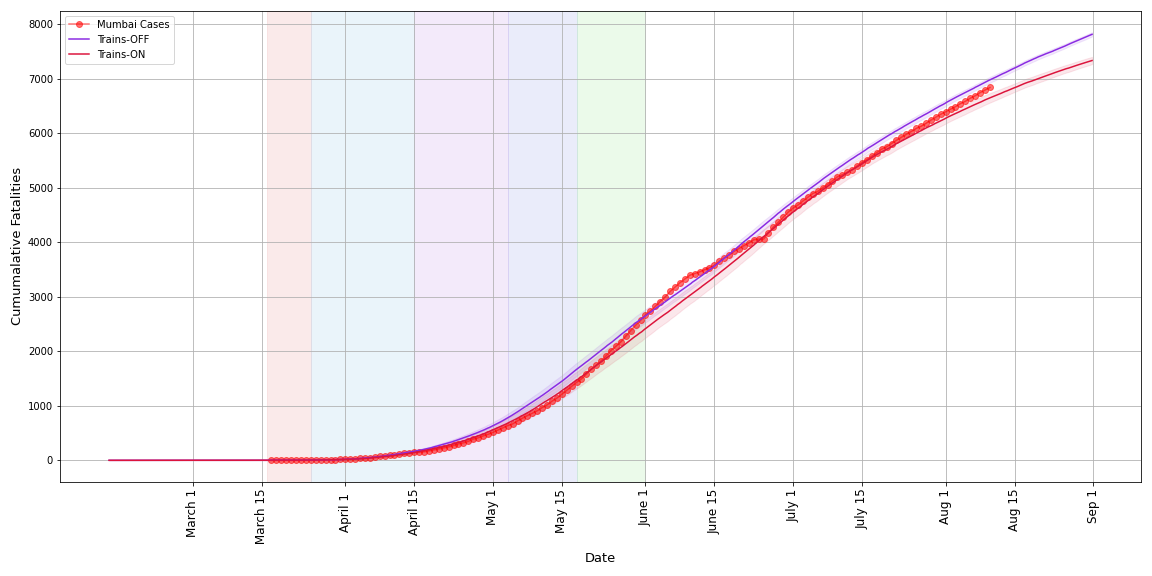}
\caption{\small Case study C, subsection~\ref{subsec:casestudyC}: Mumbai cumulative fatalities estimation.}
\label{fig:Mumbai-c-cumulative-fatalities}
\end{figure}

\begin{figure}
\centering
\includegraphics[scale=0.4]{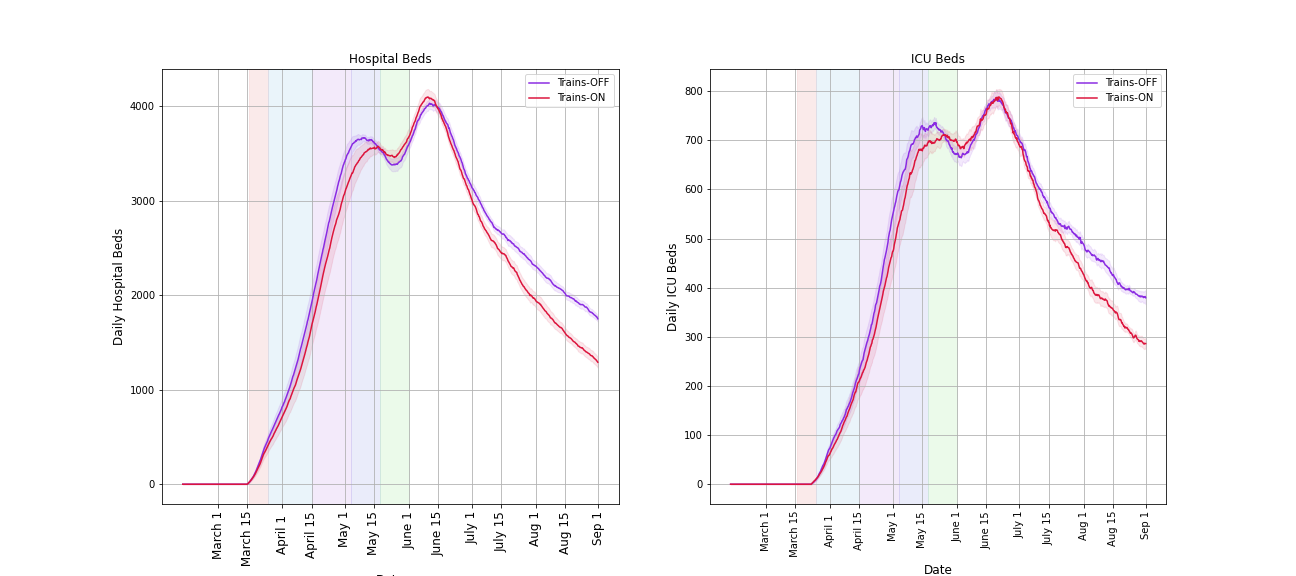}
\caption{\small Case study C, subsection~\ref{subsec:casestudyC}: Mumbai hospital beds estimation.}
\label{fig:Mumbai-c-hospital-beds}
\end{figure}

We now study the impact of opening suburban trains in  Mumbai. Table~\ref{tab:Mumbai-interventions} shows the timeline of various restrictions implemented in Mumbai starting from 01\,March\,2020. Suburban trains were not under operation in Mumbai during 15\,March -- 31\,May\,2020. As suburban trains are a key mode of daily commute in Mumbai, we compare the situation in Table~\ref{tab:Mumbai-interventions} under two scenarios:
\begin{itemize}
\item Trains-ON: Suburban trains are operational starting 01 June 2020 in a phased manner, similar to the opening of workplaces in a phased manner as indicated in Table~\ref{tab:Mumbai-interventions},
\item Trains-OFF: Suburban trains are not operational throughout.
\end{itemize}
As indicated in Table~\ref{tab:Mumbai-interventions},  we assume a compliance factor of 60\% in non-slums and 40\% in slums. We plot our results in Figures~\ref{fig:Mumbai-c-daily-cases}-\ref{fig:Mumbai-c-hospital-beds}.

\begin{itemize}
\item From the plots, we see that the phased opening of suburban trains starting 01 June 2020 gives a marginal increase in the number of cases, fatalities and hospital beds compared to the Trains-OFF scenario. This suggests that trains can be operated  with enforcement of strict physical distancing (by operating at reduced passenger loads\footnote{Physical distancing under normal passenger loads in Mumbai locals is not possible given the large number of commuters per train.}) and compulsory wearing of face masks.
\item Although we match the daily fatalities\footnote{We use a corrected version of the reported number of daily fatalities from Brihanmumbai Municipal Corporation (BMC). The initial reported daily fatalities curve from BMC had a very large peak at 16 June 2020. The corrected data adjusts the daily fatalities curve until 15 June 2020 so that the peak on 16 June 2020 gets re-distributed to the previous days in a suitable way.} curve very well, we over-predict the daily number of cases. We believe that this is due to the limitation on the testing capacity on the ground. Because of this, the test results of many people arrive late and cases get reported with a certain delay. It is also worth mentioning that, although we overpredict the daily number of cases, we correctly capture the growth rate of the daily number of cases as well as the cumulative number of cases.
\end{itemize}

\FloatBarrier

\subsection{Case study D: Soft ward containment versus neighbourhood containment}
\label{subsec:casestudyD}

We study the impact of two containment strategies for Bengaluru: soft ward containment (i.e., linearly-varying mobility control that turns an open ward into a locked ward when the number of hospitalised cases become 0.1\% of the ward’s population; in the latter locked scenario, only 25\% mobility is allowed for essential services, see Figure~\ref{fig:soft-containment}) and neighbourhood containment (i.e., when an individual is hospitalised,  everyone  living in a 100m surrounding area undergoes home quarantine). Soft ward containment is a more feasible strategy than strict ward containment since the average ward population in Bengaluru is about 62,000. As the number of hospitalised cases in the ward increases, more public health wardens could be deployed and help reduce mobility and interaction in the ward.

\begin{figure}
\centering
\includegraphics[scale=0.4]{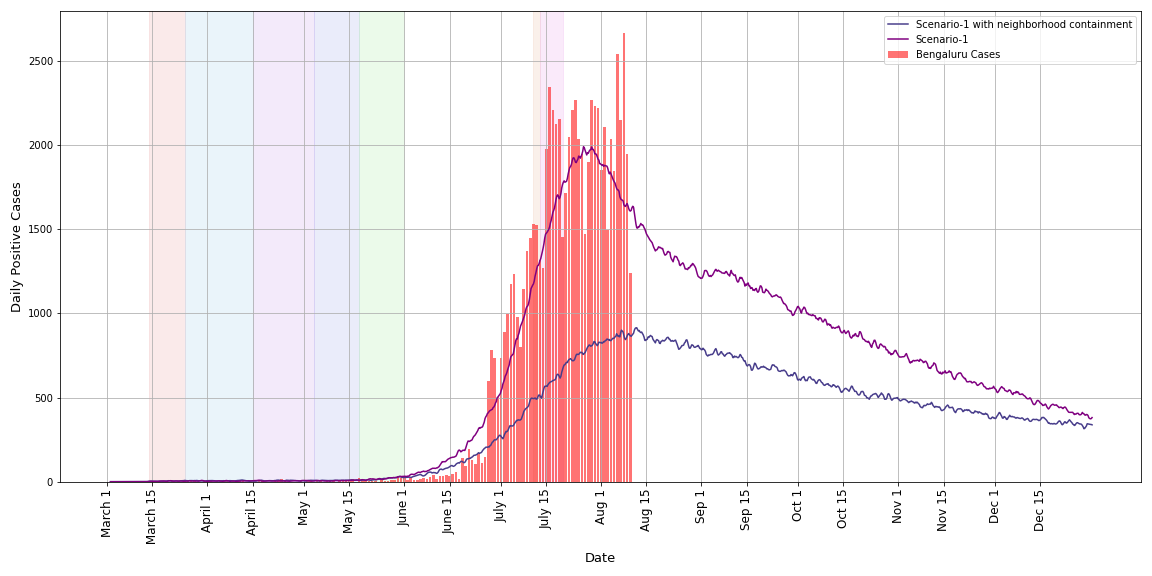}
\caption{\small Case study D, subsection~\ref{subsec:casestudyD}: Bengaluru daily cases estimation.}
\label{fig:Bengaluru-d-daily-cases}
\end{figure}

\begin{figure}
\centering
\includegraphics[scale=0.4]{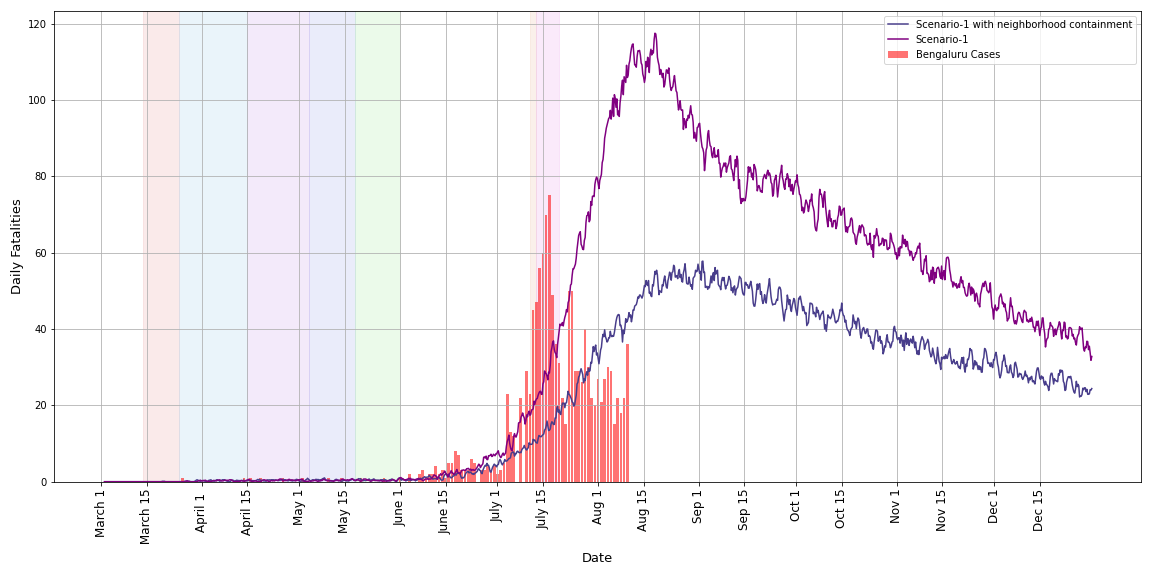}
\caption{\small Case study D, subsection~\ref{subsec:casestudyD}: Bengaluru daily fatalities estimation.}
\label{fig:Bengaluru-d-daily-fatalities}
\end{figure}

\begin{figure}
\centering
\includegraphics[scale=0.4]{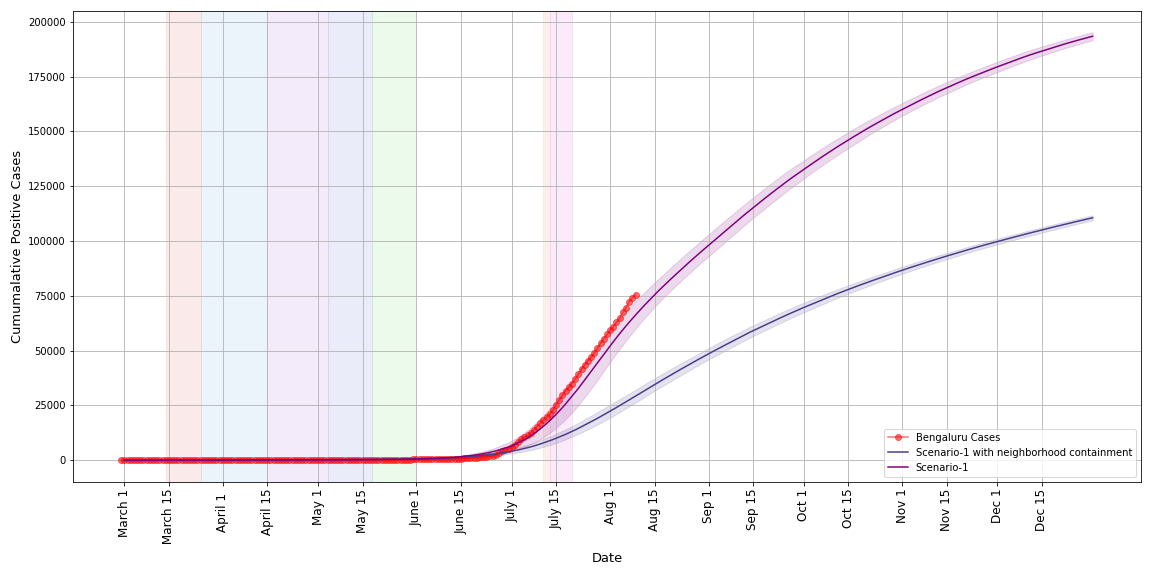}
\caption{\small Case study D, subsection~\ref{subsec:casestudyD}: Bengaluru cumulative cases estimation.}
\label{fig:Bengaluru-d-cumulative-cases}
\end{figure}

\begin{figure}
\centering
\includegraphics[scale=0.4]{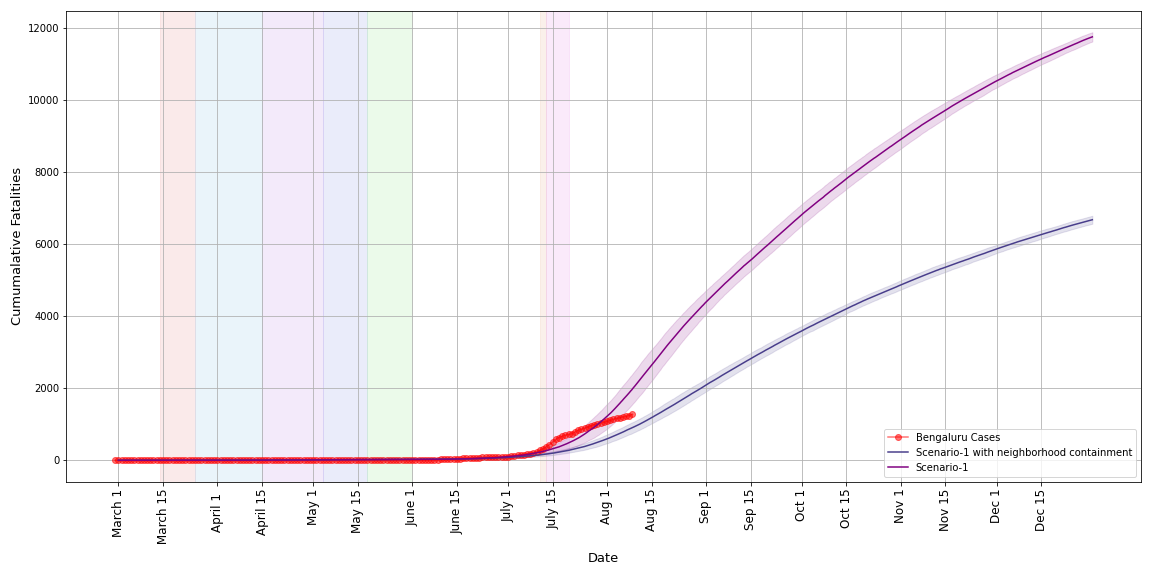}
\caption{\small Case study D, subsection~\ref{subsec:casestudyD}: Bengaluru cumulative fatalities estimation.}
\label{fig:Bengaluru-d-cumulative-fatalities}
\end{figure}

\begin{figure}
\centering
\includegraphics[scale=0.4]{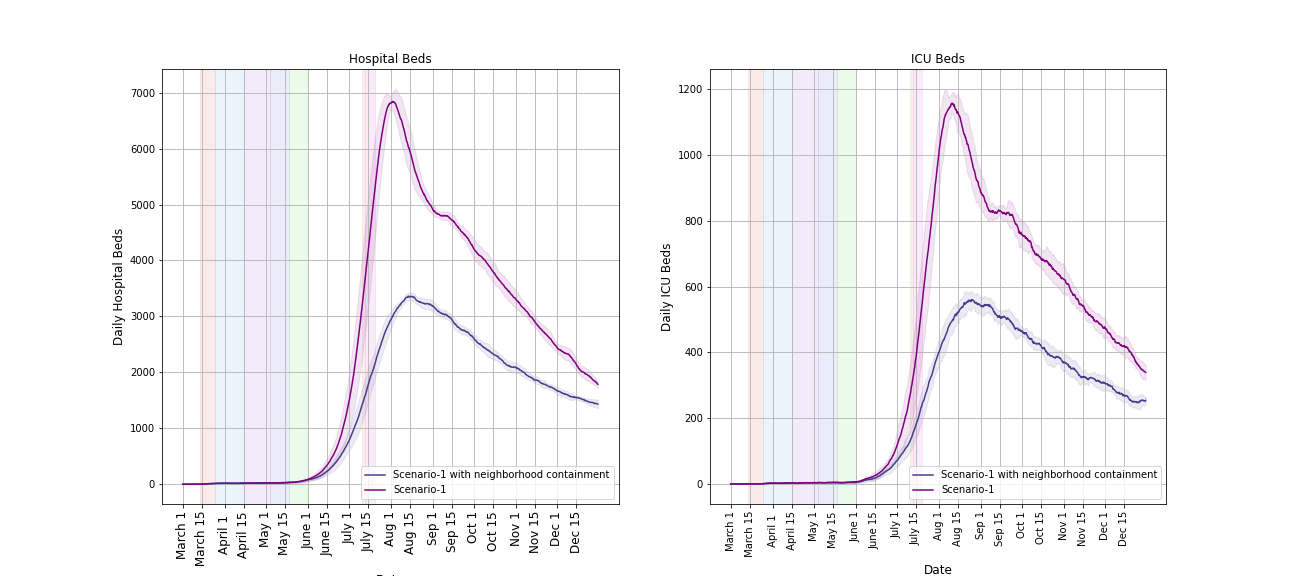}
\caption{\small Case study D, subsection~\ref{subsec:casestudyD}: Bengaluru hospital beds estimation.}
\label{fig:Bengaluru-d-hospital-beds}
\end{figure}

In Figures~\ref{fig:Bengaluru-d-daily-cases}-\ref{fig:Bengaluru-d-hospital-beds}, we plot these two scenarios. We observe that neighbourhood containment is more effective than soft ward containment, in terms of cases and fatalities.

\FloatBarrier

\subsection{Case Study E: Soft ward containment at various levels in Mumbai}
\label{subsec:casestudyE}

To compare various levels of strictness with which policies are enforced, we now consider the opening scenario indicated in Table~\ref{tab:Mumbai-interventions} and vary the containment leakage to see how this affects the numbers. Containment leakage stands for the level of activity that is allowed in a ward under containment. A strict enforcement would not allow more than 10\% of the normal activity. The case of `no enforcement' results in activity at 100\% of the original level; this corresponds to no adaptation in containment as a function of the number of hospitalisation cases. We explore various values of containment leakage and plot results for 10\%, 25\%, 50\% and 100\%.

\begin{figure}
\centering
\includegraphics[scale=0.4]{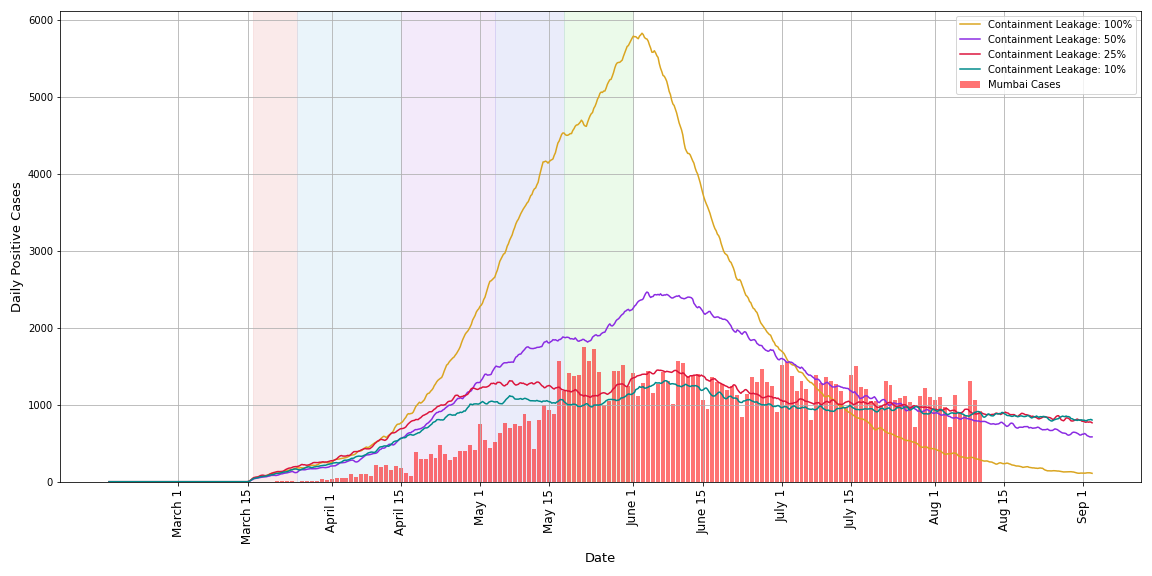}
\caption{\small Case study E, subsection~\ref{subsec:casestudyE}: Mumbai daily cases estimation.}
\label{fig:Mumbai-e-daily-cases}
\end{figure}

\begin{figure}
\centering
\includegraphics[scale=0.4]{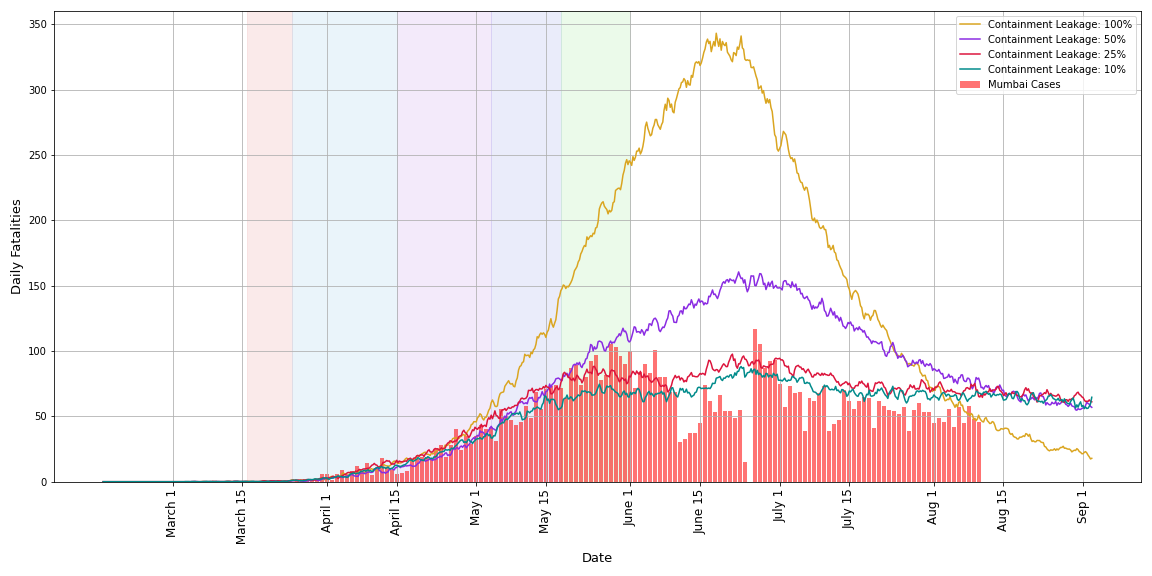}
\caption{\small Case study E, subsection~\ref{subsec:casestudyE}: Mumbai daily fatalities estimation.}
\label{fig:Mumbai-e-daily-fatalities}
\end{figure}

\begin{figure}
\centering
\includegraphics[scale=0.4]{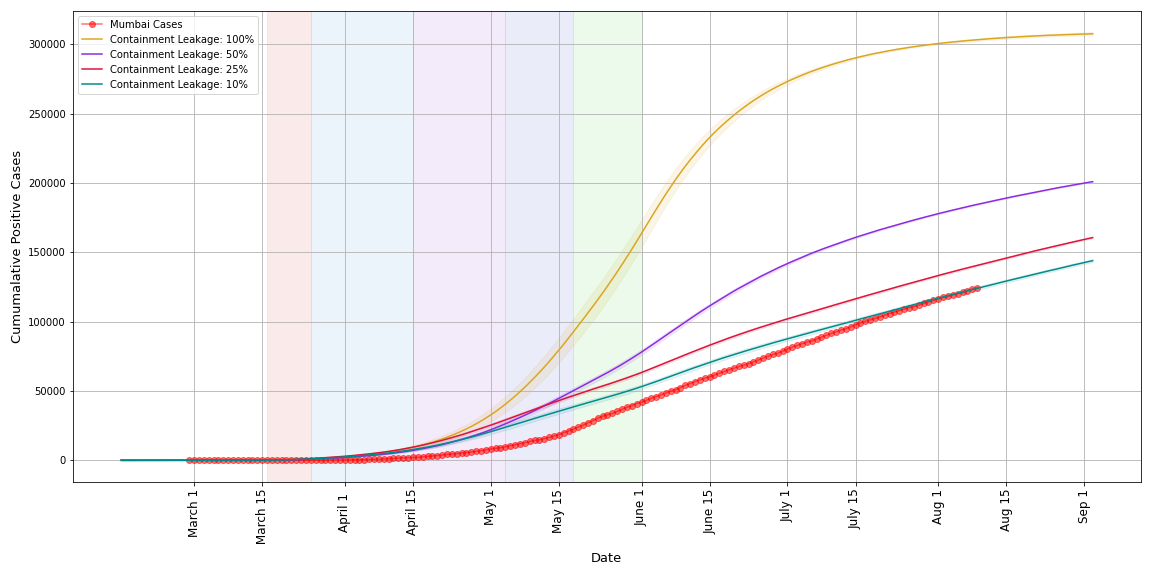}
\caption{\small Case study E, subsection~\ref{subsec:casestudyE}: Mumbai cumulative cases estimation.}
\label{fig:Mumbai-e-cumulative-cases}
\end{figure}

\begin{figure}
\centering
\includegraphics[scale=0.4]{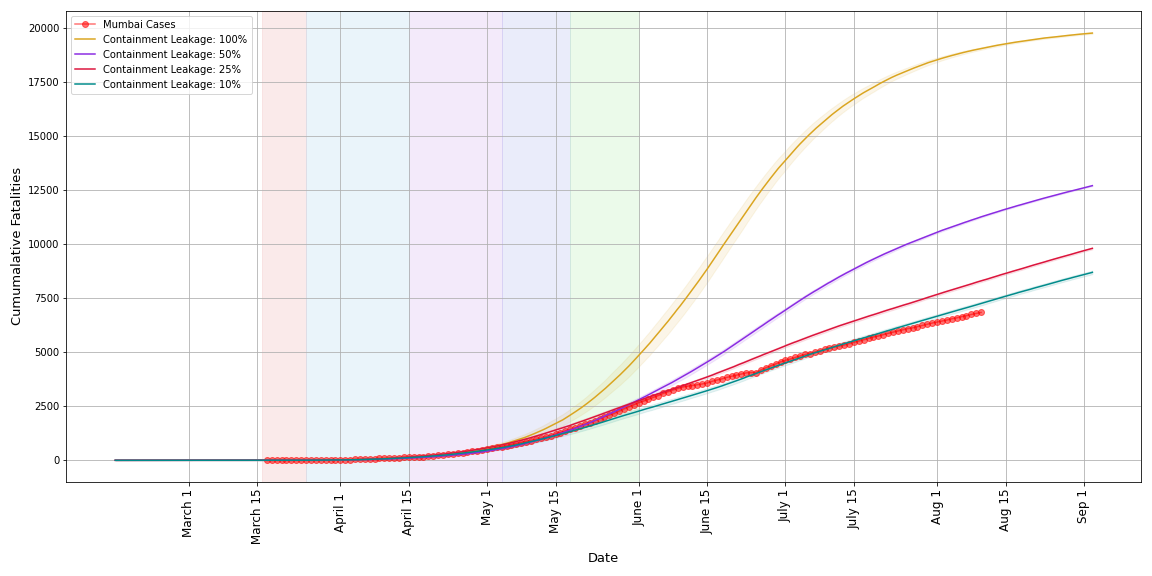}
\caption{\small Case study E, subsection~\ref{subsec:casestudyE}: Mumbai cumulative fatalities estimation.}
\label{fig:Mumbai-e-cumulative-fatalities}
\end{figure}

\begin{figure}
\centering
\includegraphics[scale=0.4]{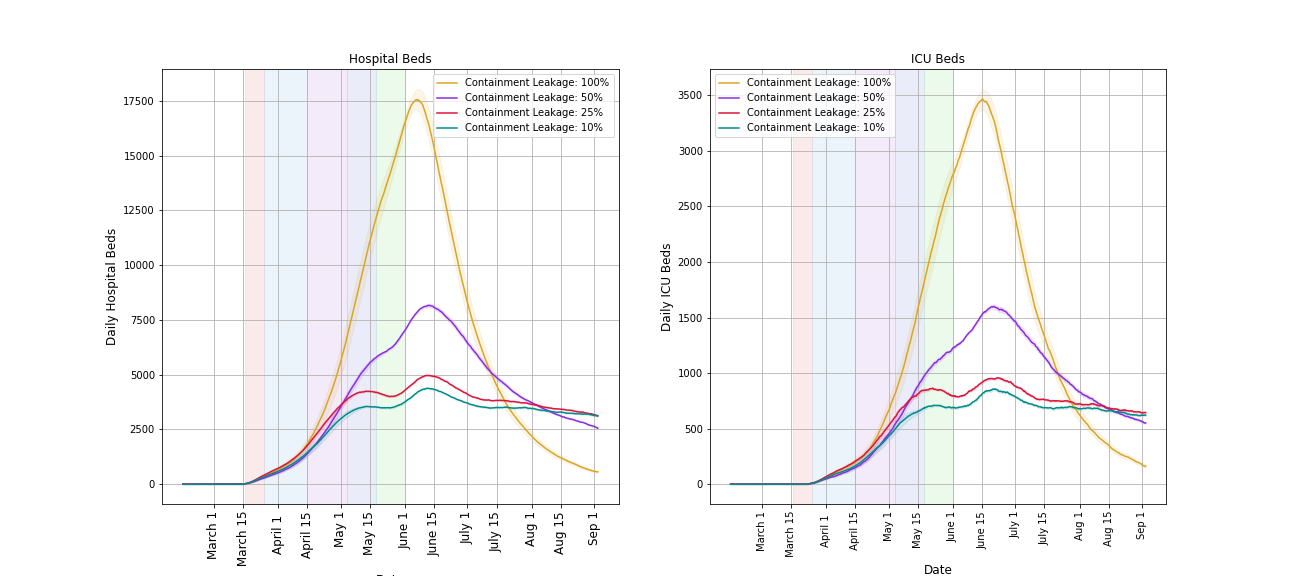}
\caption{\small Case study E, subsection~\ref{subsec:casestudyE}: Mumbai hospital beds estimation.}
\label{fig:Mumbai-e-hospital-beds}
\end{figure}

Figures~\ref{fig:Mumbai-e-daily-cases}-\ref{fig:Mumbai-e-hospital-beds}
represent the (simulated) number of daily cases, cumulative
cases, daily fatalities, cumulative fatalities and daily hospital bed estimates, respectively, for the varying containment leakages. These plots demonstrate how an effective containment policy (even as low as
50\%) can significantly reduce the number of cases, fatalities and hospital beds.

An interesting observation is that, for the 25\% leakage case, our simulator matches the linear growth trend of the daily and cumulative number of cases as well as fatalities. The linearity is a likely consequence of the specific soft ward containment policy, triggered by the hospitalisation cases, i.e., mobility is reduced linearly with the number of hospitalisations in that ward. It is not clear to what extent differential equation models can capture such linear trends.

\FloatBarrier

\subsection{Case Study F: Bengaluru with schools/colleges open from 01 September 2020}
\label{subsec:casestudyF}

\begin{figure}
\centering
\includegraphics[scale=0.4]{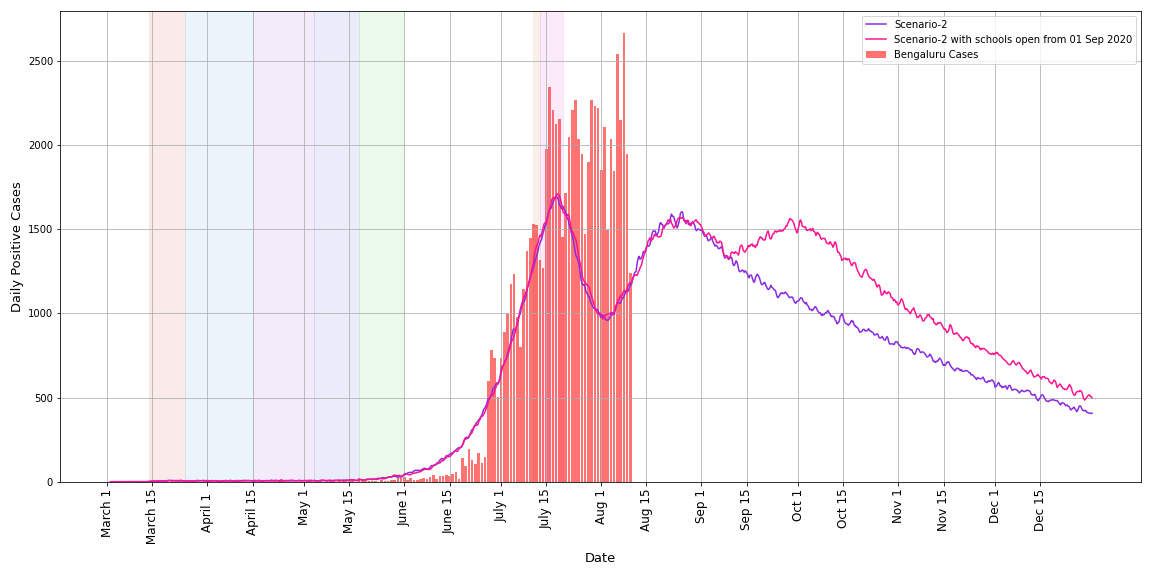}
\caption{\small Case study F, subsection~\ref{subsec:casestudyF}: Bengaluru daily cases estimation.}
\label{fig:Bengaluru-f-daily-cases}
\end{figure}

\begin{figure}
\centering
\includegraphics[scale=0.4]{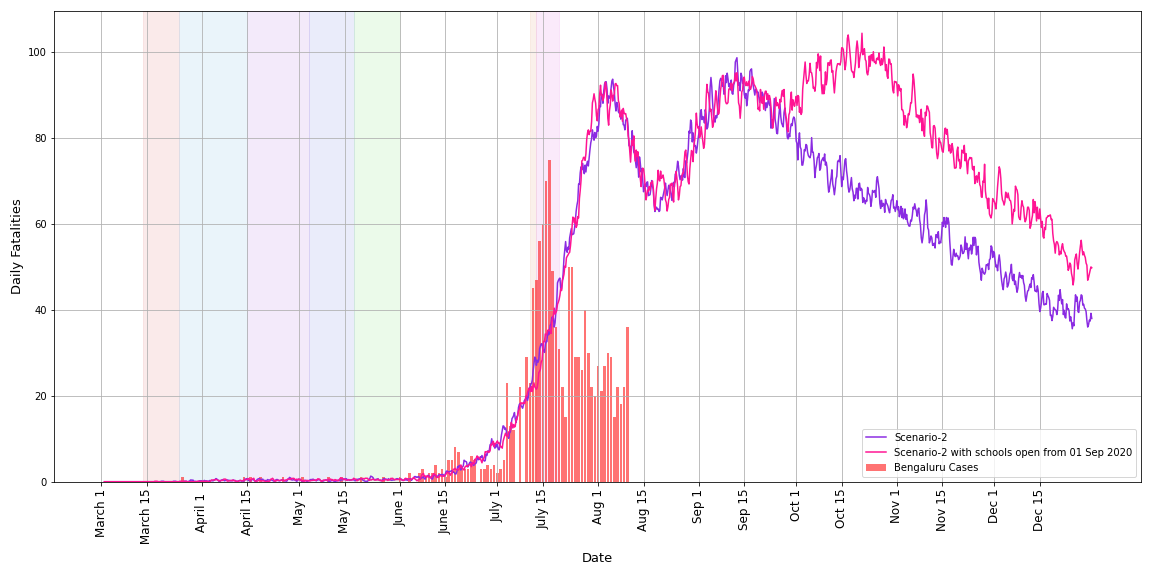}
\caption{\small Case study F, subsection~\ref{subsec:casestudyF}: Bengaluru daily fatalities estimation.}
\label{fig:Bengaluru-f-daily-fatalities}
\end{figure}

\begin{figure}
\centering
\includegraphics[scale=0.4]{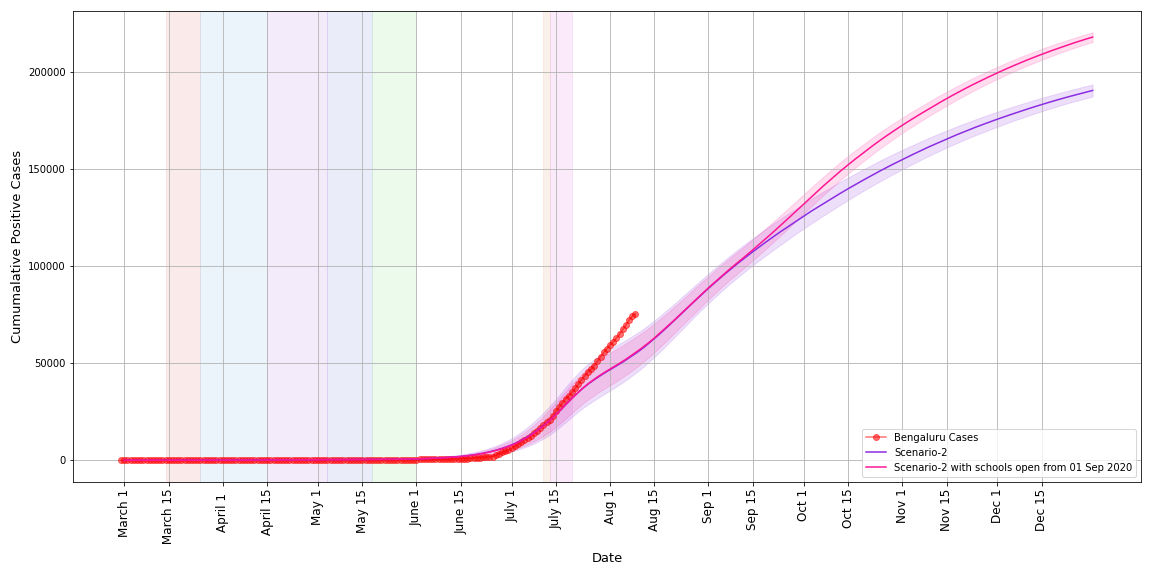}
\caption{\small Case study F, subsection~\ref{subsec:casestudyF}: Bengaluru cumulative cases estimation.}
\label{fig:Bengaluru-f-cumulative-cases}
\end{figure}

\begin{figure}
\centering
\includegraphics[scale=0.4]{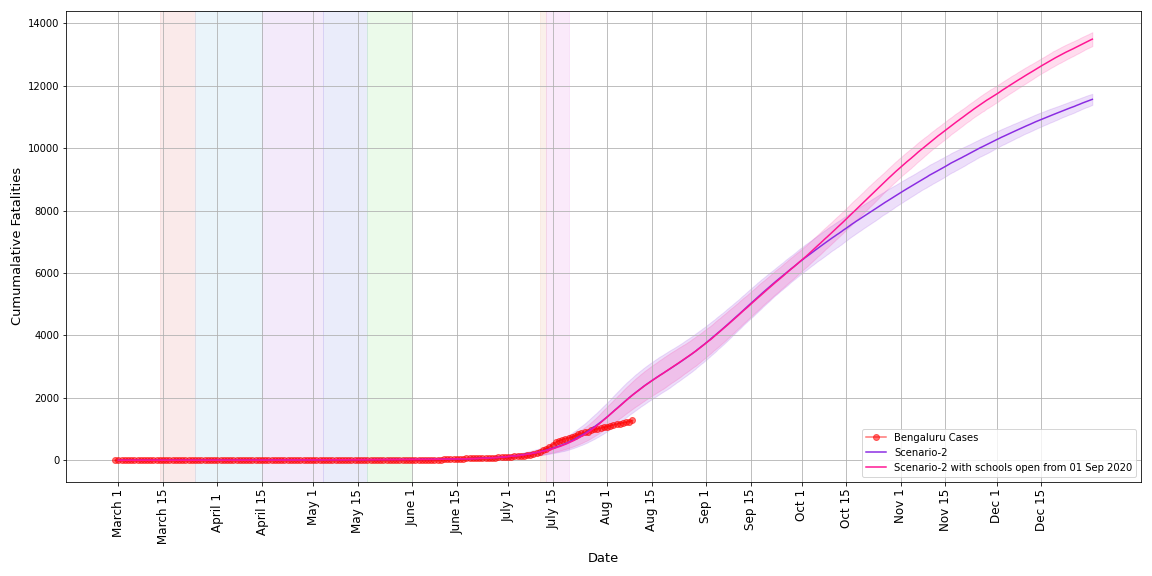}
\caption{\small Case study F, subsection~\ref{subsec:casestudyF}: Bengaluru cumulative fatalities estimation.}
\label{fig:Bengaluru-f-cumulative-fatalities}
\end{figure}

\begin{figure}
\centering
\includegraphics[scale=0.4]{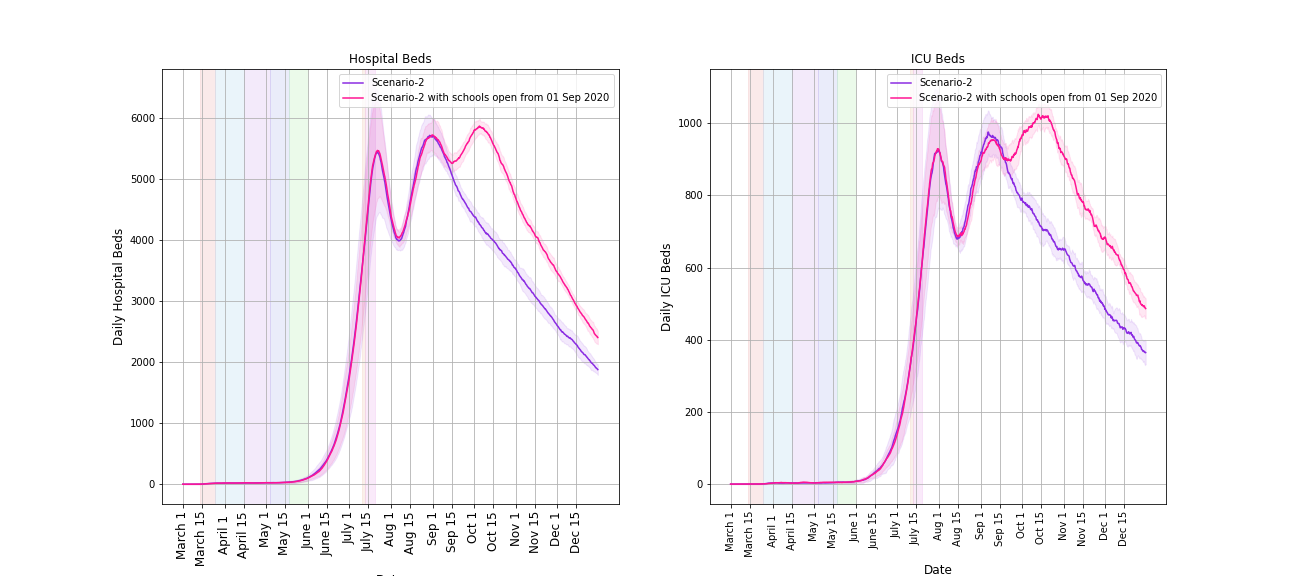}
\caption{\small Case study F, subsection~\ref{subsec:casestudyF}: Bengaluru hospital beds estimation.}
\label{fig:Bengaluru-f-hospital-beds}
\end{figure}

We now study the impact of opening schools. In Figures~\ref{fig:Bengaluru-f-daily-cases}--\ref{fig:Bengaluru-f-hospital-beds}, we compare the following two scenarios:
\begin{itemize}
\item Schools-closed: The present scenario in Bengaluru, i.e., Scenario-2 in Table~\ref{tab:Bengaluru-interventions},
\item Schools-open: Scenario-2 in Table~\ref{tab:Bengaluru-interventions} with schools open from 01 September 2020.
\end{itemize}
As expected, both these scenarios follow the same trend until about mid-September, after which the disease spread increases in the latter. Around early November, we observe a between 10-15\% increase in the cumulative number of cases and the cumulative number of fatalities due to the opening of schools. This could be weighed with other factors such as the capacity of our healthcare system to handle the rise, the impact of mental health of students due to extended closures, etc., while arriving at a decision on whether schools can be opened from 01 September 2020. The proportion of additional children and adults affected due to the opening of the schools is still under study.

\FloatBarrier

\section{Simulator}
\label{sec:Simulator}
\subsection{City generation}
The first step in our agent-based model is to model a synthetic city that respects the demographics of the city that we want to study. Our city generator uses the following data as input:
\begin{itemize}
  \item Geo-spatial data that provides information on the wards of a city (components) along with boundaries. (If this is not available, one could feed in ward centre locations and ward areas).
  \item Population in each ward, with break up on those living in high density and low density areas.
  \item Age distribution in the population.
  \item Household size distribution (in high and low density areas) and some information on the age composition of the houses (e.g., generation gaps, etc.)
  \item The number of employed individuals in the city.
  \item Distribution of the number of students in schools and colleges.
  \item Distribution of the workplace sizes.
  \item Distribution of commute distances.
  \item Origin-destination densities that quantify movement patterns within the city.
\end{itemize}

\vspace*{.1in}

Taking the above data into account, individuals, households, workplaces, schools, transport spaces, and community spaces are instantiated. Individuals are then assigned to households, workplaces or schools, transport and community spaces, see Figure~\ref{fig:abm-art} for a schematic representation. The algorithms for the assignments do a coarse matching. The matching may be refined as better data becomes available.

\begin{figure}[!ht]
\centering
\includegraphics[scale=0.5]{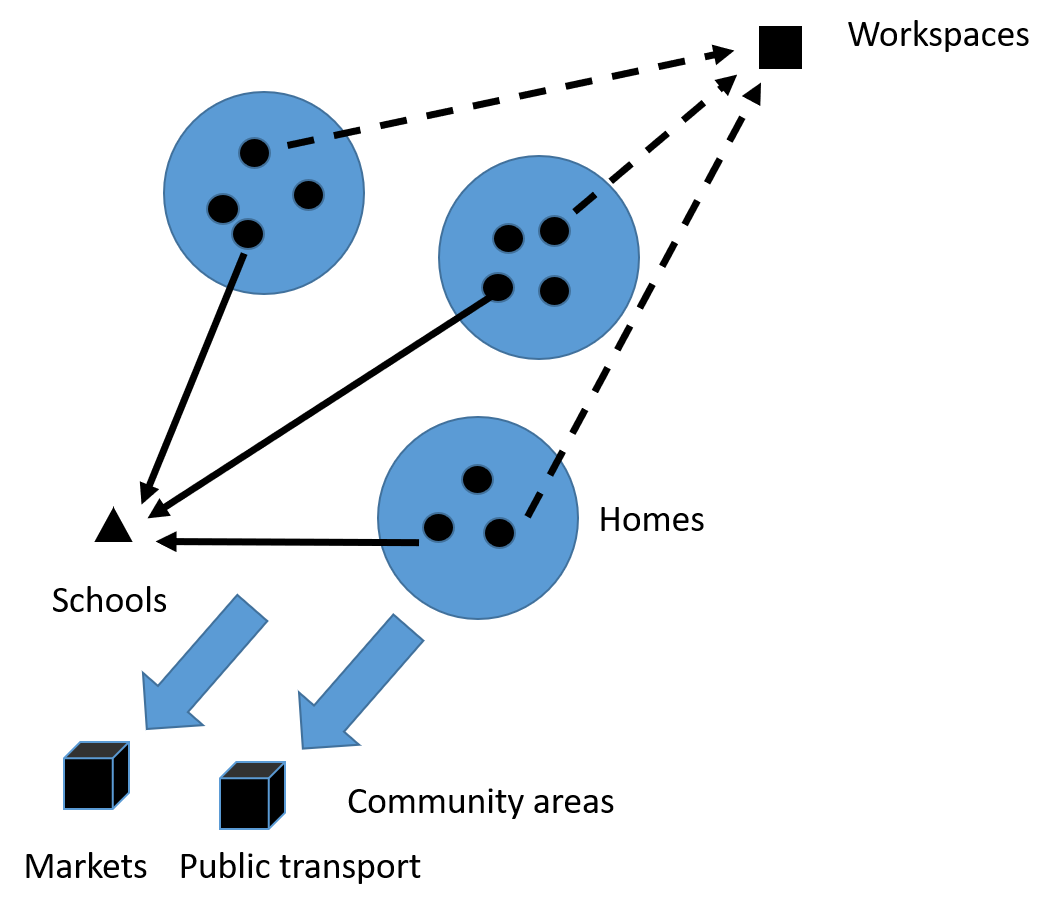}
\caption{\small Various interaction spaces, solid circles inside homes indicate individuals.}
\label{fig:interaction-spaces}
\end{figure}

\begin{figure}[!ht]
\centering
\includegraphics[scale=0.5]{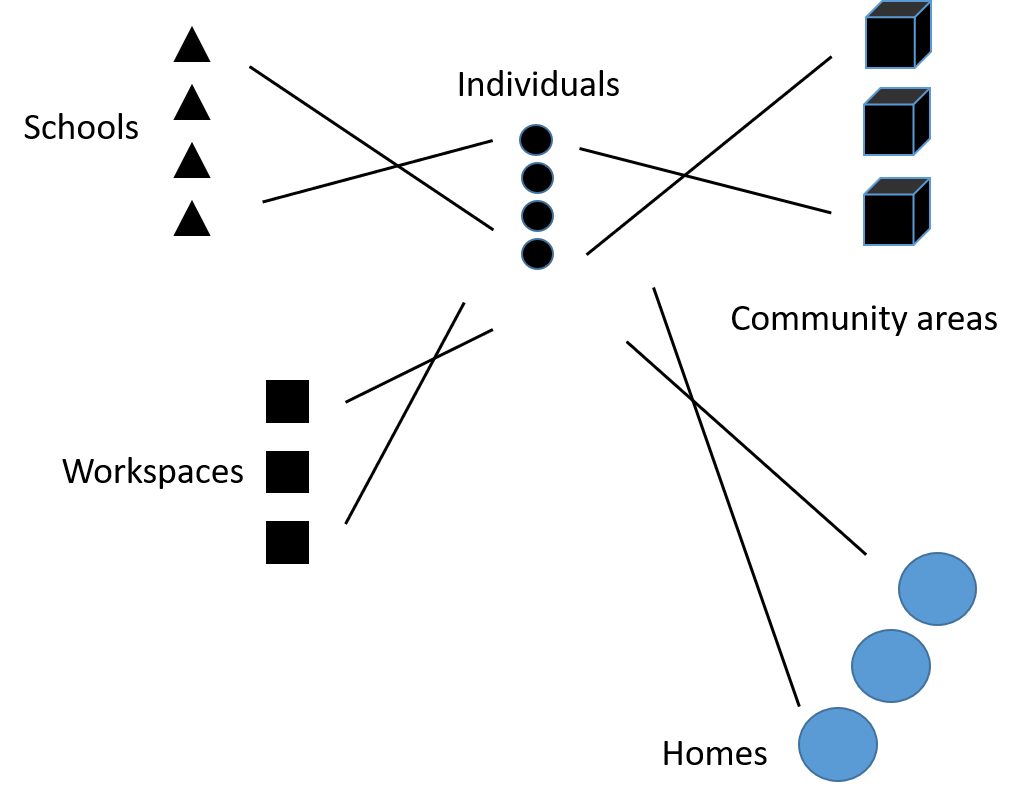}
\caption{\small Bipartite graph abstraction of interaction spaces.}
\label{fig:interaction-spaces-bipartite}
\end{figure}

The interaction spaces -- households, workplaces or schools, transport and community spaces -- reflect different social networks and transmission happens along their edges. There is interaction among these graphs because the nodes are common across the graphs, see Figure~\ref{fig:interaction-spaces} for various interaction spaces and Figure~\ref{fig:interaction-spaces-bipartite} for a bipartite graph abstractions of these interaction spaces. An individual of school-going age who is exposed to the infection at school may expose others at home. This reflects an interaction between the school graph and the household graph. Similarly other graphs interact.

We now describe how individuals are assigned to interaction spaces.

\begin{figure}
\begin{subfigure}{.5\textwidth}
  \centering
\includegraphics[scale=0.15]{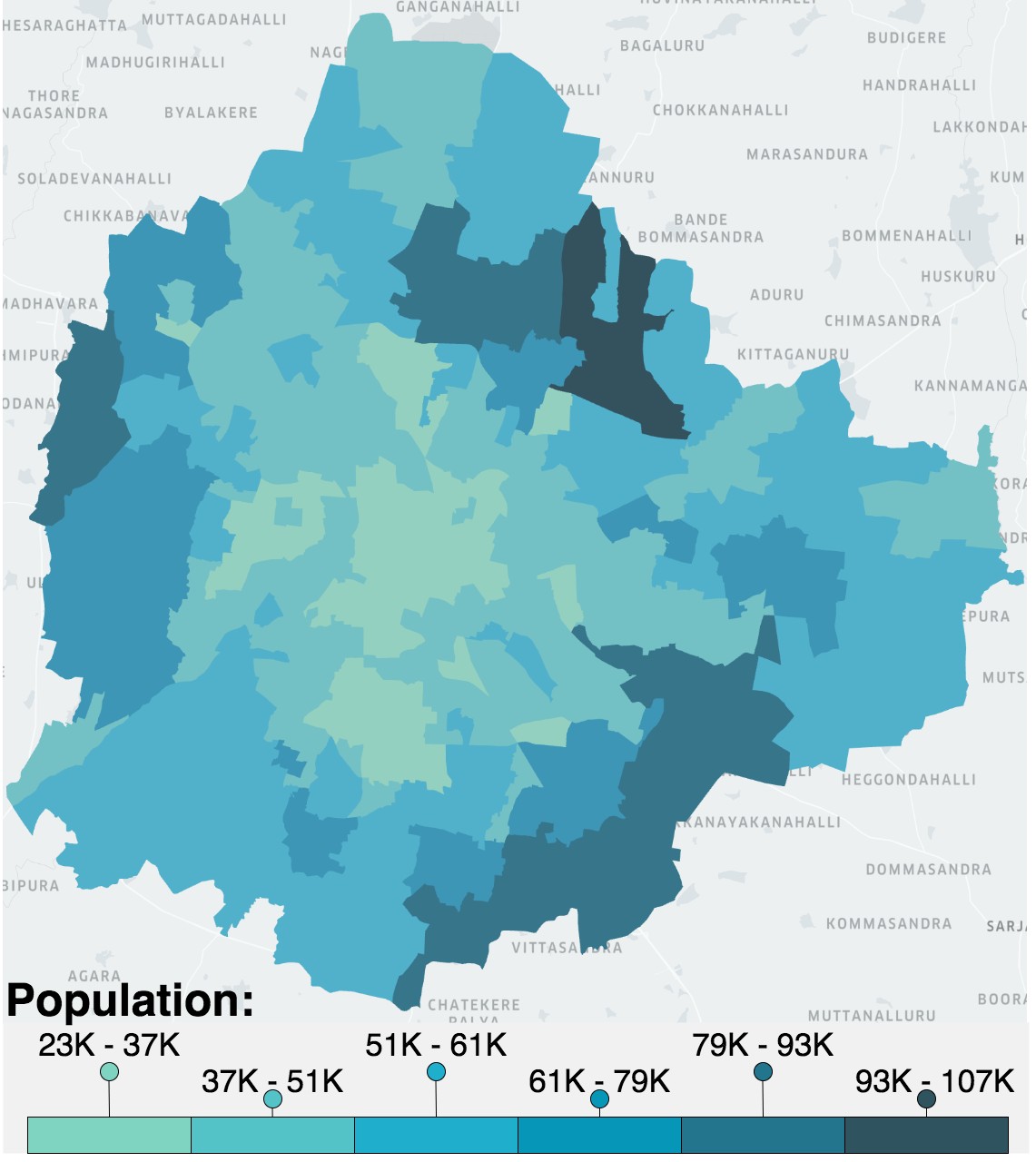}
\caption{\small Bengaluru.}
\label{fig:Bengaluru-population}
\end{subfigure}
\begin{subfigure}{.5\textwidth}
  \centering
\includegraphics[scale=0.15]{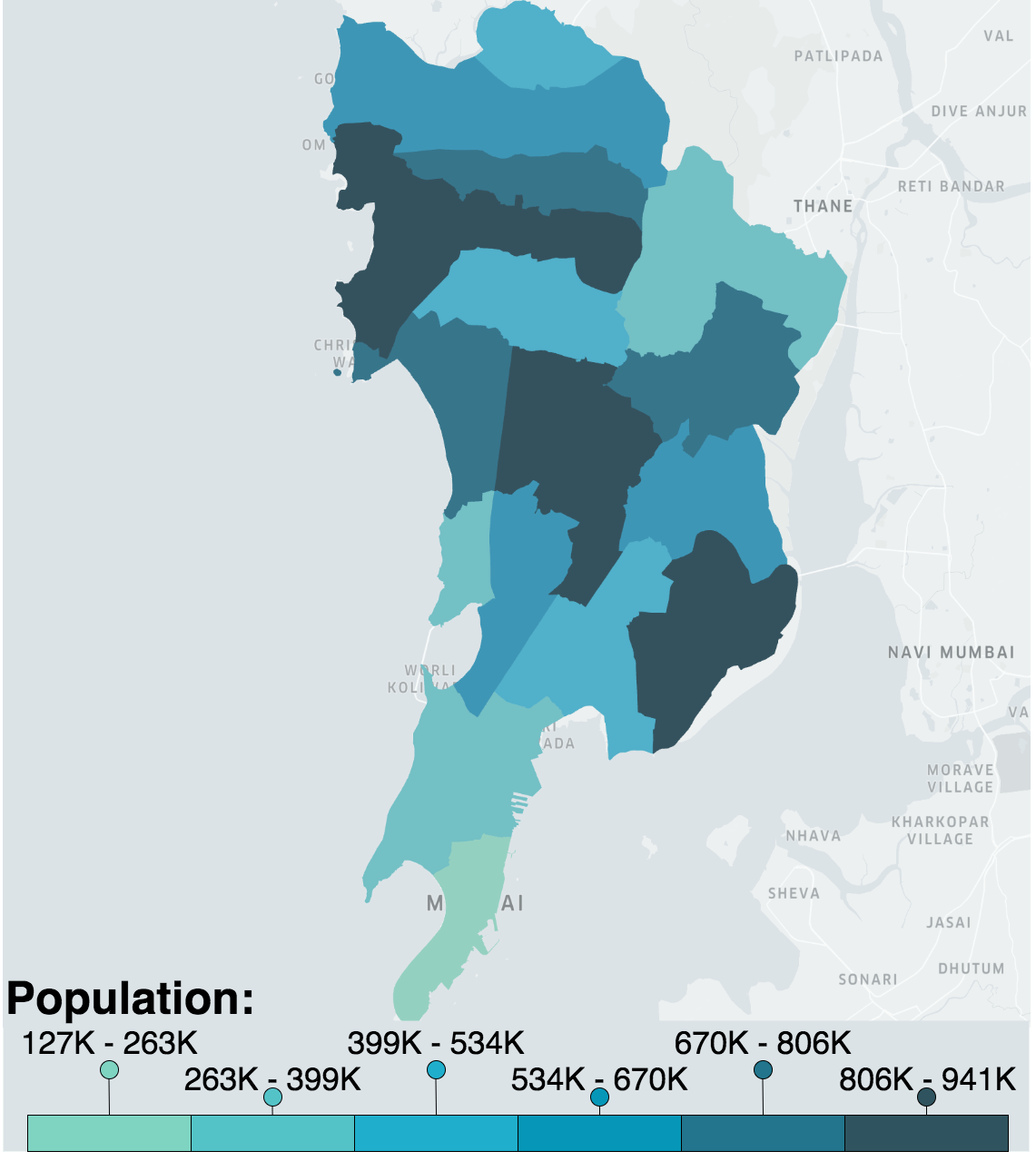}
\caption{\small Mumbai.}
\label{fig:Mumbai-population}
\end{subfigure}
\caption{\small Population density maps of Bengaluru and Mumbai.}
\label{fig:population-density-maps}
\end{figure}

{\em Individuals and households}: $N$ individuals are instantiated and ages are sampled according to the age distribution in the population. Households (based on the $N$ and the mean number per household) are then instantiated and assigned a random number of individuals sampled according to the distribution of household sizes. An assignment of individuals to households is then done to match, to the extent possible, the generational structure in typical households. The households are then assigned to wards so that the total number of individuals in the ward is in proportion to population density in the ward, taken from census data. A population density map is given in Figure~\ref{fig:population-density-maps}(\subref{fig:Bengaluru-population}) for Bengaluru and in Figure~\ref{fig:population-density-maps}(\subref{fig:Mumbai-population}) for Mumbai. The generational gap, household distribution, and age distribution patterns are assumed to be uniform across the wards in the city. Each household in a ward is then assigned a random location in the ward, and all individuals associated with the household are assigned the same geo-location as the household.

Based on the age and the unemployment fraction, each individual is either a student or a worker or neither.

\vspace*{.1in}

{\em Assignment of schools}: Children of school-going ages 5-14 and a certain fraction of the population aged 15-19 are assigned to schools. These are taken to be students. The remaining fraction of the population aged 15-19 and a certain fraction of the population aged 20-59, based on information on the employed fraction\footnote{The unemployed fraction in Bengaluru, from the census data, is just over 50\%, even after taking into account employment in the unorganised sector. Similar is the case with Mumbai. This may have some bearing on the epidemic spread.}, are all classified as workers and are assigned workplaces. The rest of the population (nonstudent, unemployed) is not assigned to either schools or workplaces.

In past works, given the structure of educational institutions elsewhere, educational institutions have been divided into primary schools, secondary schools, higher secondary schools, and universities. The norm in Indian urban areas is that schools handle primary to higher secondary students and then colleges handle undergraduates. We view all such entities as schools.

We assign students to schools on a ward-by-ward basis. In each ward, we have a certain number of students. We pick a school size from a given school size distribution and instantiate a school of this size and place it randomly in that ward. Students who live in that ward are picked randomly and assigned to this school until that school is filled to its capacity. We repeat this procedure until all students in that ward gets assigned to a school, and then we repeat this procedure for all wards. This procedure could lead to at most one school per ward whose capacity is more than its sampled capacity.

\vspace*{.1in}

{\em Assignment of workplaces}: Workplace interactions can enable the spread of an epidemic. In principle, Bengaluru's and Mumbai's land-use data could be used to locate office spaces. The assignment of individuals to workplaces is done in two steps. In the first step, for each individual who goes to work, we decide the ward where his/her office is located. This assignment of a ``working ward" is based on an Origin-Destination (OD) matrix. An OD matrix is a square matrix whose number of rows equals the number of wards, and its $(i, j)$th entry tells us the fraction of people who travel from ward $i$ to ward $j$ for work. In the second step, for each ward, we sample a workplace size from a workplace size distribution, create a workplace of this capacity and place it uniformly-at-random in that ward. We then randomly assign individuals who work in that ward to this workplace. Similar to assignment of schools, we continue to create workplaces in this ward until every individual working in that ward gets assigned to a workplace, and we repeat this procedure for all wards. For Bengaluru, the OD matrix is obtained from  the regional travel model used for Bengaluru. For Mumbai, based on the ``zone to zone'' travel data from \cite{worldbank}, an origin-destination matrix was extrapolated based on the population of each ward.

The above assignments could be improved further in later versions of this simulator.

\vspace*{.1in}

{\em Community spaces}: Community spaces include day care centres, clinics, hospitals, shops, markets, banks, movie halls, marriage halls, malls, eateries, food messes, dining areas and restaurants, public transit entities like bus stops, metro stops, bus terminals, train stations, airports, etc. While we hope to return to model a few of the important ones explicitly at a later time, we proceed along the route taken by \cite{ferguson2005strategies} with two modifications.

In our current implementation, each individual sees one community that is personalised to the individual's location and age and one transport space personalised to the individual's commute distance. For ease of implementation, the personalisation of the community space is based on ward-level common local communities and a distance-kernel based weighting. The personalisation of the transport space is based on commute distance. Details are given in Section~\ref{subsec:model-of-infection-spread}.

{\em Age-stratified interaction}: The interactions across these communities could be age-stratified. This may be informed by social networks studies, for e.g., as in \cite{prem2017projecting} which has been used in a recent compartmentalised SEIR model \cite{singh2020age}.

{\em Smaller subnetworks}: We create smaller subnetworks in workplaces, schools and communities, and associate certain number people to these smaller networks with the interpretation that people in a smaller subnetwork have high contact rate among them compared to the others. In some more detail, we create ``project" networks at each workplace consisting of people in that workplace having closer interaction, a ``class" network in each school consisting of students of the same age, a random community network among people in a given ward to model daily random interactions, and a neighbourhood subnetwork among people living in a $178m \times 178m$ square \footnote{This dimension $178m$ comes from an approximate ``squaring of a circle" of $100m$ radius}. These subnetworks are later used for identifying and testing/quarantining individuals based on a contact tracing protocol.

\begin{figure}
\begin{subfigure}{.5\textwidth}
  \centering
  \includegraphics[width=.7\linewidth]{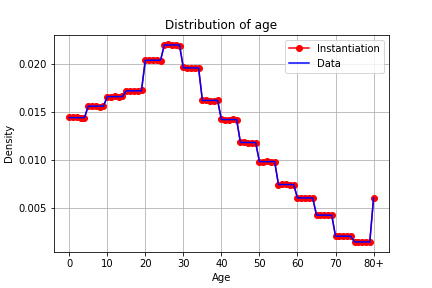}
  \caption{\small Bengaluru -- Age distribution}
  \label{fig:validation-blr-age}
\end{subfigure}
\begin{subfigure}{.5\textwidth}
  \centering
  \includegraphics[width=.7\linewidth]{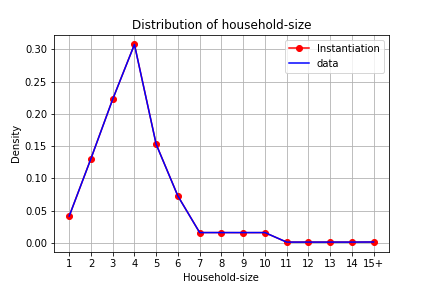}
  \caption{\small Bengaluru -- Household size distribution}
  \label{fig:validation-blr-hs}
\end{subfigure}

\begin{subfigure}{.5\textwidth}
  \centering
  \includegraphics[width=.7\linewidth]{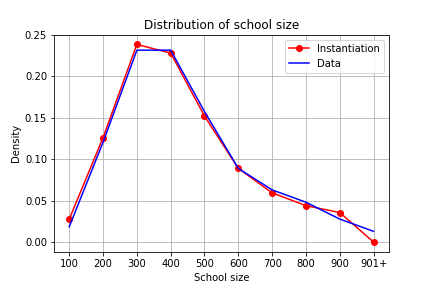}
  \caption{\small Bengaluru -- School size distribution}
  \label{fig:validation-blr-ss}
\end{subfigure}
\begin{subfigure}{.5\textwidth}
  \centering
  \includegraphics[width=.7\linewidth]{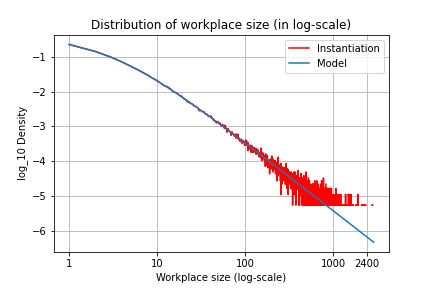}
  \caption{\small Bengaluru -- Workplace size distribution}
  \label{fig:validation-blr-ws}
\end{subfigure}

\begin{subfigure}{.5\textwidth}
  \centering
  \includegraphics[width=.7\linewidth]{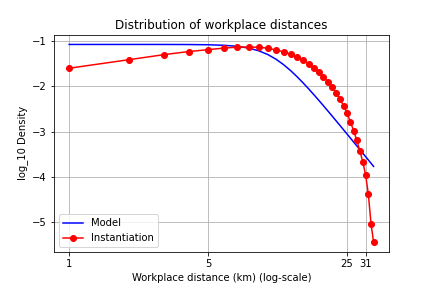}
  \caption{\small Bengaluru -- Commuter distance distribution}
  \label{fig:validation-blr-wd}
\end{subfigure}
\begin{subfigure}{.5\textwidth}
  \centering
  \includegraphics[width=.7\linewidth]{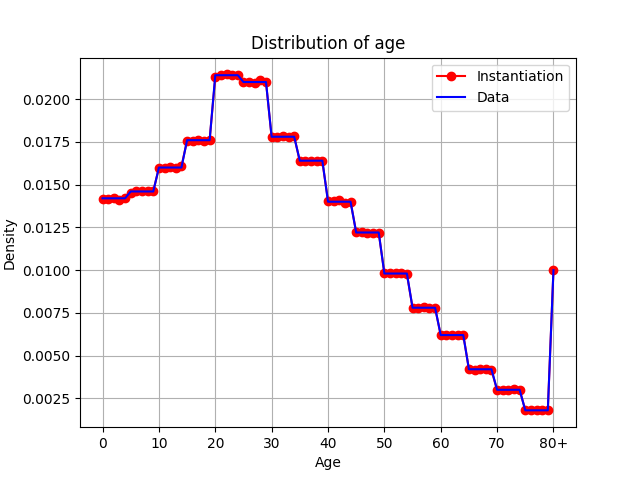}
  \caption{\small Mumbai -- Age distribution}
  \label{fig:validation-mum-age}
\end{subfigure}

\begin{subfigure}{.5\textwidth}
  \centering
  \includegraphics[width=.7\linewidth]{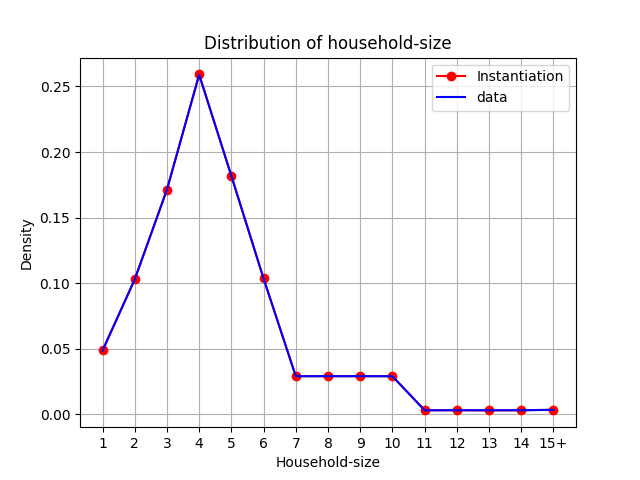}
  \caption{\small Mumbai -- Household size distribution}
  \label{fig:validation-mum-hs}
\end{subfigure}
\begin{subfigure}{.5\textwidth}
  \centering
  \includegraphics[width=.7\linewidth]{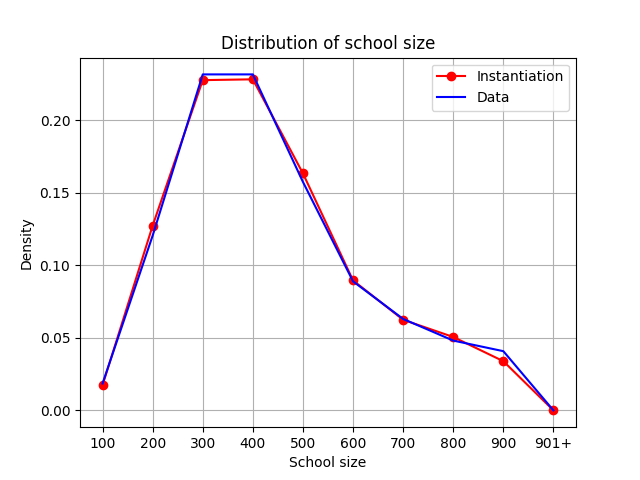}
  \caption{\small Mumbai -- School size distribution}
  \label{fig:validation-mum-ss}
\end{subfigure}

\begin{subfigure}{.5\textwidth}
  \centering
  \includegraphics[width=.7\linewidth]{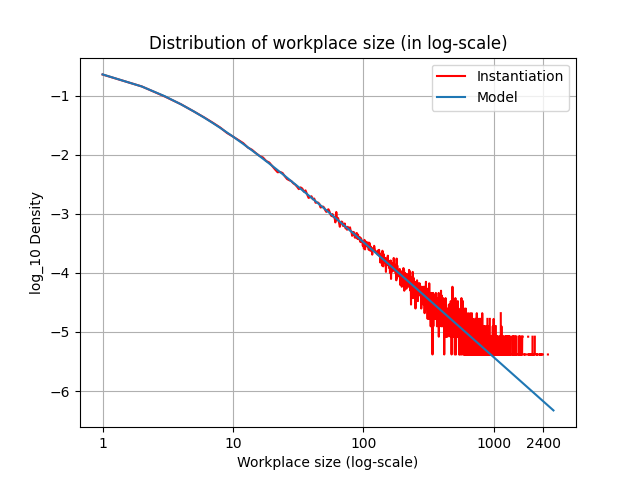}
  \caption{\small Mumbai -- Workplace size distribution}
  \label{fig:validation-mum-ws}
\end{subfigure}
\begin{subfigure}{.5\textwidth}
  \centering
  \includegraphics[width=.7\linewidth]{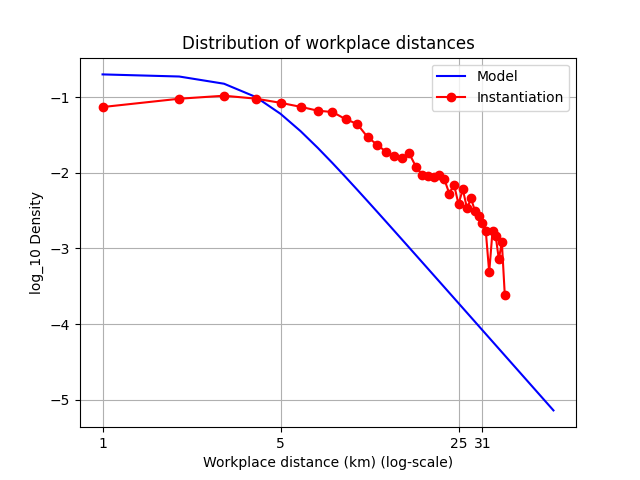}
  \caption{\small Mumbai -- Commuter distance distribution}
  \label{fig:validation-mum-wd}
\end{subfigure}

\caption{\small Validation of our synthetic Bengaluru and Mumbai. Figures~\ref{fig:validation-plots}(\subref{fig:validation-blr-age})--\ref{fig:validation-plots}(\subref{fig:validation-blr-wd}) show the validation plots for Bengaluru and Figures~\ref{fig:validation-plots}(\subref{fig:validation-mum-age})--\ref{fig:validation-plots}(\subref{fig:validation-mum-wd}) show the validation plots for Mumbai.}
\label{fig:validation-plots}
\end{figure}

The output of all the above is our synthetic city on which infection spreads. Figure~\ref{fig:validation-plots} provides an indication of how close our synthetic city is to the true city in terms of the indicated statistics.

\subsection{Disease progression}
We have used a simplified model of COVID-19 progression, based on descriptions in \cite{verity2020estimates} and \cite{ferguson2020report}. This will need updating as we get India specific data.

\begin{figure}[ht]
\centering
\includegraphics[scale=0.5]{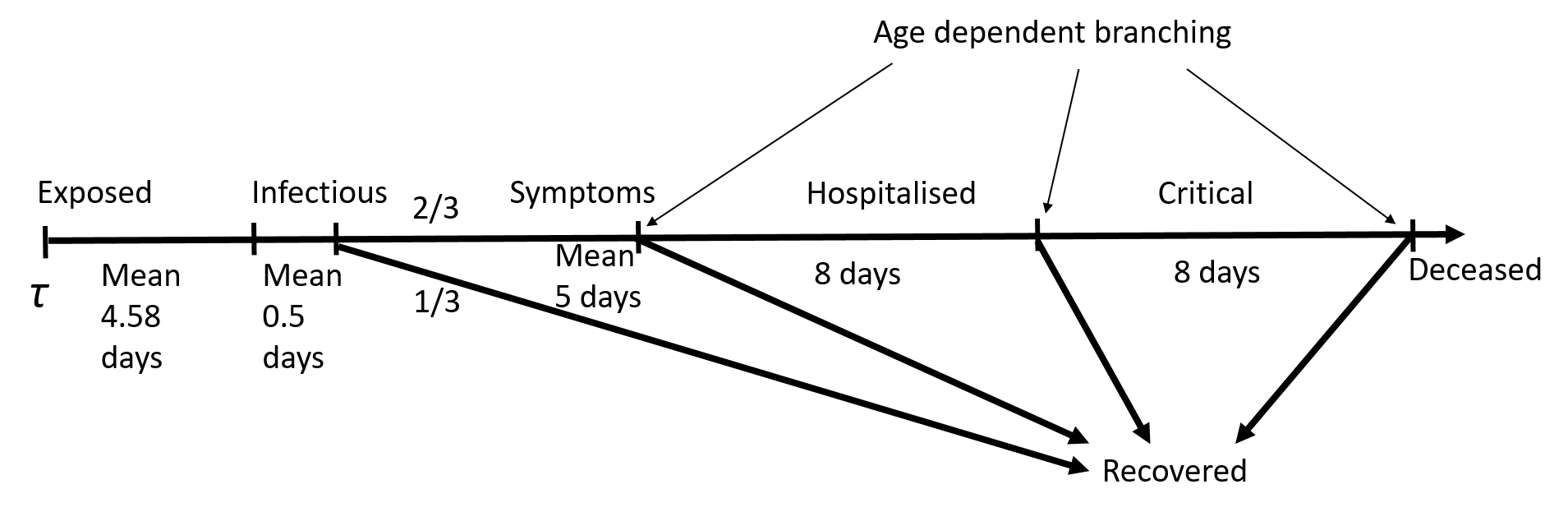}
\caption{\small A simplified model of COVID-19 progression.}
\label{fig:diseaseProgression}
\end{figure}

An individual may have one of the following states, see  Figure~\ref{fig:diseaseProgression}: susceptible, exposed, infective (pre-symptomatic or asymptomatic), recovered, symptomatic, hospitalised, critical, or deceased.

\vspace*{.1in}

We assume that initially the entire population is susceptible to the infection. Let $\tau$ denote the time at which an individual is exposed to the virus, see Figure~\ref{fig:diseaseProgression}. The incubation period is random with the Gamma distribution of shape 2 and scale 2.29; the mean incubation period is then 4.58 days (4.6 days in \cite{ferguson2020report} and 4.58 in \cite{prem2020effect}). Individuals are infectious for an exponentially distributed period of mean duration 0.5 of a day. This covers both presymptomatic transmission and possible asymptomatic transmission. We assume that a third of the patients recover, these are the asymptomatic patients; the remaining two-third develop symptoms. Estimates of the number of asymptomatic patients vary from 0.2 to 0.6. Though we have explored other asymptomatic fractions, we restrict attention here to 1/3. Symptomatic patients are assumed to be 1.5 times more infectious during the symptomatic period than during the pre-symptomatic but infective stage. Individuals either recover or move to the hospital after a random duration that is exponentially distributed with a mean of 5 days\footnote{This needs to be updated based on hospitalisation guidelines.}. The probability that an individual recovers depends on the individual's age\footnote{It is possible to add comorbidities - diabetes, hypertension, etc. -- in addition to age. Mortality and prognosis appear to depend heavily on comorbidities. We leave it for the future.}. It is also assumed that recovered individuals are no longer infective nor susceptible to a second infection. While hospitalised individuals may continue to be infectious, they are assumed to be sufficiently isolated, and hence do not further contribute to the spread of the infection. Further progression of hospitalised individuals to critical care is mainly for assessing the need for hospital beds, intensive care unit (ICU) beds, critical care equipments, etc. This will need to be adapted to our local hospital protocol.

Let us reiterate. Once a susceptible individual has been exposed, the trajectory in Figure~\ref{fig:diseaseProgression} takes over for that individual. Further progressions are (in our current implementation) only based on the agent's age.

\begin{table}[ht]
\caption{Hospitalisation estimates taken from \cite{ferguson2020report}, based on studies in \cite{verity2020estimates}.}
\label{tab:demand}
\centering
\begin{tabular}{|l|c|c|c|}
 \hline
 Age-group & \% symptomatic cases       & \% hospitalised cases & \% critical cases  \\
 (years)   & requiring hospitalisation  & requiring critical care & deceased \\
 [0.5ex]
 \hline
 \hline
 0 to 9   & 0.1\% & 5.0\% & 40\% \\
 10 to 19 & 0.3\% & 5.0\% & 40\% \\
 20 to 29 & 1.2\% & 5.0\% & 50\% \\
 30 to 39 & 3.2\% & 05.0\% & 50\% \\
 40 to 49 & 4.9\% & 6.3\% & 50\% \\
 50 to 59 & 10.2\% & 12.2\% & 50\% \\
 60 to 69 & 16.6\% & 27.4\% & 50\% \\
 70 to 79 & 24.3\% & 43.2\% & 50\% \\
 80+      & 27.3\% & 70.9\% & 50\% \\
 \hline
\end{tabular}
\end{table}

\subsection{Disease spread}
\label{subsec:model-of-infection-spread}
At each time $t$, an infection rate $\lambda_n(t)$ is computed for each individual $n$ based on the prevailing conditions. In the time duration $\Delta t$ following time $t$, each susceptible individual moves to the exposed state with probability $1 - \exp\{ - \lambda_n(t) \cdot \Delta t\}$, independently of all other events. Other transitions are as per the disease progression described earlier. Time is then updated to $t + \Delta t$, the conditions are then updated to reflect the new exposures, changes to infectiousness, hospitalisations, recoveries, contact tracing, quarantines, tests, test outcomes, etc., during the period $t$ to $t + \Delta t$. The process outlined at the beginning of this paragraph is repeated until the end of the simulation. $\Delta t$ was taken to be 6 hours in our simulator and is configurable.

A city of $N$ individuals, $H$ households, $S$ schools, $W$ workplaces, one community space (comprising $C$ wards), one transport space, and associations of individuals to these entities is the starting point for the infection spread simulator. Infection spread is then implemented as follows.

An individual $n$ can transmit the virus in the infective (pre-symptomatic or asymptomatic stage) or in the symptomatic stage. At time $t$, this is indicated as $I_n(t) = 1$ when infective and otherwise $I_n(t) = 0$ otherwise.

Additionally, each individual has two other parameters: a severity variable $C_n$ and a relative infectiousness variable $\rho_n$, see \cite{ferguson2005strategies}. Both bring in heterogeneity to the model. Severity $C_n = 1$ if the individual suffers from a severe infection and $C_n = 0$ otherwise; this is sampled at 50\% probability independently of all other events. Infectiousness $\rho_n$ is a random variable that is Gamma distributed with shape 0.25 and scale 4 (so the mean is 1). The severity variable captures severity-related absenteeism at school/workplace, associated decrease of infection spread at school/workplace, and the increase of infection spread at home.

If the individual gets exposed at time $\tau_n$, a relative infection-stage-related infectiousness is taken to be $\kappa(t - \tau_n)$ at time $t$. For the disease progression described in the previous section, this is 1 in the presymptomatic and asymptomatic stages, 1.5 in the symptomatic, hospitalised, and critical stages, and 0 in the other stages.

To describe the infection spread at transport spaces, let $\mathcal{T}(n) =1$ if agent $n$ uses public transport and let $\mathcal{T}(n) =0$ otherwise. Let $A_{n,t}=0$ if at time $t$ agent $n$ is either (i) compliant and under quarantine, (ii) hospitalised, (iii) critical, or (iv) dead, and let $A_{n,t} = 1$ if none of the above is true and agent $n$ attends office at time $t$. We model the
effectiveness of masks by reducing the ability of an infectious individual to transmit the infection by 20\%, if a mask is worn (see~\cite{mask-respiratory2009,maskuse-influenza2010,mask-respiratoryvirus2020,chu2020physical}); let $M_n=0.8$ if agent $n$ wears a face mask in public transport and $M_n=1$ otherwise.

Let $\beta_h$, $\beta_s$, $\beta_w$, $\beta_T$, $\beta_c$, $\beta_h^*, \beta_s^*, \beta_w^*$ and $\beta_c^*$ denote the transmission coefficients at home, school, workplace, transport, community spaces, neighbourhood network, class network, project network and random community network, respectively. These can be viewed as scaled contact rates with members in the household, school, workplace, community, neighbourhood, class, project and random community, respectively. More precisely, these are the {\em expected number of eventful (infection spreading) contact opportunities} in each of these interaction spaces. It accounts for the combined effect of frequency of meetings and the probability of infection spread during each meeting.

For a susceptible individual, the rate of transmission is governed by the sum of product of contact rate $\beta$ and infectiousness in all the interactions spaces. To model infectiousness, we consider three scenarios.\\
{\em Interactions without age-stratification}: This is the simplest model where interactions within each network is assumed to be homogeneous. A susceptible individual $n$ (who belongs to home $h(n)$, school $s(n)$, workplace $w(n)$, transport space $T(n)$, and community space $c(n)$) sees the following infection rate at time $t$:

\begin{eqnarray}
\lambda_n(t)
  & = & \sum_{n' : h(n') = h(n)} \frac{1}{n_{h(n)}^{\alpha}} \cdot I_{n'}(t) \beta_{h} \kappa(t - \tau_{n'}) \rho_{n'} (1 + C_{n'}(\omega - 1))\nonumber \\
  & & +~ \sum_{n' : s(n') = s(n)} \frac{1}{n_{s(n)}} \cdot I_{n'}(t) \beta_{s} \kappa(t - \tau_{n'}) \rho_{n'} (1 + C_{n'}(\omega \psi_s(t - \tau_n) - 1)) \nonumber \\
  & & +~ \sum_{n' : w(n') = w(n)} \frac{1}{n_{w(n)}} \cdot I_{n'}(t) \beta_{w} \kappa(t - \tau_{n'}) \rho_{n'} (1 + C_{n'}(\omega \psi_w(t - \tau_n) - 1)) \nonumber \\
  & & +~ \frac{\sum_{n^\prime : \mathcal{T}(n^\prime) =1} A_{n^\prime,t} }{\sum_{n^\prime} \mathcal{T}(n^\prime)}   \times \sum_{n':T(n')=T(n)} \left( \frac{d_{n', w(n')}  I_{n'}(t) \beta_{T} M_{n^\prime}}{\sum_{n':T(n')=T(n)} d_{n',w(n')}} \right) \nonumber \\
  & & +~ \frac{\zeta(a_n) \cdot f(d_{n,c})}
  {\sum_{c'}f(d_{c,c'})} \sum_{c'} f(d_{c,c'})
  h_{c,c'}(t) \label{eqn:rate}
\end{eqnarray}
where
\begin{equation}
 \label{eqn:community-interaction}
 h_{c,c'}(t) = \left( \frac{\sum_{n': c(n') = c'} f(d_{n',c(n')}) \cdot \zeta(a_{n'}) \cdot I_{n'}(t) \beta_{c} r_c \kappa(t - \tau_{n'}) \rho_{n'} (1 + C_{n'}(\omega - 1))}{\sum_{n'} f(d_{n',c(n')})}  \right)
\end{equation}

The expression \eqref{eqn:rate} can be viewed as the rate at which the susceptible individual $n$ contracts the infection at time $t$. Each of the components on the right-hand side indicates the rate from home, school, workplace, transport space, and community. The additional quantities, over and above what we have already described, are as follows.

The parameter $\alpha$ determines how household transmission rate scales with household size, a crowding-at-household factor. It increases the propensity to spread the infection by a factor $n^{1-\alpha}$. We have taken $\alpha = 0.8$, see \cite{ferguson2005strategies}.

A common parameter $\omega$ indicates how a severely infected person affects a susceptible one, as will be clear from below. (This is to be tuned at a later stage and is set to 2 now).

The functions $\psi_s(\cdot)$ and $\psi_w(\cdot)$ account for absenteeism in case of a severe infection. It can be time-varying and can depend on school or workplace. We take $\psi_s(t) = 0.1$ and $\psi_w(t) = 0.5$ while infective and after one day since infectiousness. School-goers with severe infection contribute lesser to the infection spread, due to higher absenteeism, than those that go to workplaces; moreover, the absenteeism results in an increased spreading rate at home.

The function $\zeta(a)$ is the relative travel-related contact rate of an individual aged $a$. We take this to be 0.1, 0.25, 0.5, 0.75, 1, 1, 1, 1, 1, 1, 1, 1, 0.75, 0.5, 0.25, 0.1 for the various age groups in steps of 5 years, with the last one being the 80+ category.

The quantity $h_{c,c'}(t)$ represents the transmission rate from individuals in ward $c'$ to an individual in ward $c$. As above, each individual contributes in a distance-weighted way in how an individual in a ward $c'$ affects another individual in another ward $c$.

The factor $r_c$ stands for a high-density interaction multiplying factor. For Mumbai, $r_c = 2$ for some high density areas and $r_c = 1$ for the other areas. For Bengaluru $r_c = 1$ for all wards.

The function $f(\cdot)$ is a distance kernel that can be matched to the travel patterns in the city.

Finally, our choice of the infection rate from the community space is a little different from the rate specified in \cite{ferguson2005strategies}, in order to enable an efficient implementation. When the distance kernel is $f(d) = 1/(1+(d/a)^b)$ and $d \ll a$, i.e., the wards are small, then our specification is close to that indicated in \cite{ferguson2005strategies}. We take $a = 10.751$\,km and $b = 5.384$, based on a fit on data for Bengaluru.

As one can see from \eqref{eqn:rate}, we have one community space but with contributions from various wards. This enables inclusion of `containment zones' and the associated restriction of interaction across such zones, as we shall soon describe.

{\em Age-stratified interactions}: If this is enabled, the home, school, workplace and community interaction rates have an extra factor $M^h_{n,n'}, M^s_{n,n'}$, and $M^w_{n,n'}$ in the summand which accounts for age-stratified interactions. Each of these depends on $n$ and $n'$ only through the ages of agents $n$ and $n'$. The resulting contact rate for individual $n$ at time $t$ is then:

\begin{eqnarray}
\lambda_n(t)
  & = & \sum_{n' : h(n') = h(n)} \frac{M^h_{n,n'}}{n_{h(n)}^{\alpha}} \cdot I_{n'}(t) \beta_{h} \kappa(t - \tau_{n'}) \rho_{n'} (1 + C_{n'}(\omega - 1))\nonumber \\
  & & +~ \sum_{n' : s(n') = s(n)} \frac{M^s_{n,n'}}{n_{s(n)}} \cdot I_{n'}(t) \beta_{s} \kappa(t - \tau_{n'}) \rho_{n'} (1 + C_{n'}(\omega \psi_s(t - \tau_n) - 1)) \nonumber \\
  & & +~ \sum_{n' : w(n') = w(n)} \frac{M^w_{n,n'}}{n_{w(n)}} \cdot I_{n'}(t) \beta_{w} \kappa(t - \tau_{n'}) \rho_{n'} (1 + C_{n'}(\omega \psi_w(t - \tau_n) - 1)) \nonumber \\
  & & +~ \frac{\sum_{n^\prime : \mathcal{T}(n^\prime) =1} A_{n^\prime,t} }{\sum_{n^\prime} \mathcal{T}(n^\prime)}   \times \sum_{n':T(n')=T(n)} \left( \frac{d_{n', w(n')}  I_{n'}(t) \beta_{T}  M_{n^\prime}}{\sum_{n':T(n')=T(n)} d_{n',w(n')}} \right) \nonumber \\
  & & +~ \frac{\zeta(a_n) \cdot f(d_{n,c})}
  {\sum_{c'}f(d_{c,c'})} \sum_{c'} f(d_{c,c'})
  h_{c,c'}(t)
  \label{eqn:rate-age-stratified}
\end{eqnarray}
where $h_{c,c'}(t)$ is given in \eqref{eqn:community-interaction}. Computational complexity can be reduced by focusing only on the principal components of $M^h, M^s$, and $M^w$.

{\em Interactions with smaller subnetworks}: In this situation, we have additional contact rate parameters, one for each smaller subnetwork: let $\beta_h^*, \beta_s^*, \beta_w^*$ and $\beta_c^*$ denote the transmission coefficients at neighbourhood network, class network, project network and random community network respectively. Then, an agent $n$ (who belongs to neighbourhood network $\mathscr{H}(n)$, class $\mathscr{S}(n)$, project $\mathscr{W}(n)$ and random community $\mathscr{C}(n)$, in addition to home $h(n)$, school $s(n)$, workplace $w(n)$, transport space $T(n)$, and community space $c(n)$)  sees the following infection rate at time $t$:
\begin{eqnarray}
\lambda_n(t)
  & = & \sum_{n' : h(n') = h(n)} \frac{1}{n_{h(n)}^{\alpha}} \cdot I_{n'}(t) \beta_{h} \kappa(t - \tau_{n'}) \rho_{n'} (1 + C_{n'}(\omega - 1)) \nonumber \\
  & & +~ \zeta(a_{n}) \sum_{n' : \mathscr{H}(n') = \mathscr{H}(n)} \frac{1}{n_{\mathscr{H}(n)}} \cdot \zeta(a_{n'}) I_{n'}(t) \beta_h^* \kappa(t - \tau_{n'}) \rho_{n'} (1 + C_{n'}(\omega - 1)) \nonumber \\
  & & \hspace*{5cm} \mbox{(larger neighbourhood interaction)} \nonumber \\
  & & +~ \frac{\zeta(a_n) f(d_{n,c(n)})}{\sum_{n' : \mathscr{C}(n') = \mathscr{C}(n)} f(d_{n',c(n')}) } \nonumber \\
  & & \qquad \times \sum_{n' : \mathscr{C}(n') = \mathscr{C}(n)} f(d_{n',c(n')}) \zeta(a_{n'}) I_{n'}(t) \beta_c^* \kappa(t - \tau_{n'}) \rho_{n'} (1 + C_{n'}(\omega - 1)) \nonumber \\
  & & \hspace*{5cm} \mbox{(close friends' circle interaction)} \quad \nonumber \\
  & & +~ \sum_{n' : s(n') = s(n)} \frac{1}{n_{s(n)}} \cdot I_{n'}(t) \beta_{s} \kappa(t - \tau_{n'}) \rho_{n'} (1 + C_{n'}(\omega \psi_s(t - \tau_n) - 1)) \nonumber \\
  & & +~ \sum_{n' : \mathscr{S}(n') = \mathscr{S}(n)} \frac{1}{n_{\mathscr{S}(n)}} \cdot I_{n'}(t) \beta_s^* \kappa(t - \tau_{n'}) \rho_{n'} (1 + C_{n'}(\omega \psi_s(t - \tau_n) - 1))  \nonumber \\
  & & \hspace{5cm} \mbox{(class network interaction)} \nonumber \\
  & & +~ \sum_{n' : w(n') = w(n)} \frac{1}{n_{w(n)}} \cdot I_{n'}(t) \beta_{w} \kappa(t - \tau_{n'}) \rho_{n'} (1 + C_{n'}(\omega \psi_w(t - \tau_n) - 1)) \nonumber \\
  & & +~ \sum_{n' : \mathscr{W}(n') = \mathscr{W}(n)} \frac{1}{n_{\mathscr{W}(n)}} \cdot I_{n'}(t) \beta_{w}^* \kappa(t - \tau_{n'}) \rho_{n'} (1 + C_{n'}(\omega \psi_w(t - \tau_n) - 1)) \nonumber  \\
  & & \hspace{5cm} \mbox{(project network interaction)} \nonumber \\
  & & +~ \frac{\sum_{n^\prime : \mathcal{T}(n^\prime) =1} A_{n^\prime,t} }{\sum_{n^\prime} \mathcal{T}(n^\prime)}   \times \sum_{n':T(n')=T(n)} \left( \frac{d_{n', w(n')}  I_{n'}(t) \beta_{T}  M_{n^\prime}}{\sum_{n':T(n')=T(n)} d_{n',w(n')}} \right) \nonumber \\
  & & +~ \frac{\zeta(a_n) \cdot f(d_{n,c})}
  {\sum_{c'}f(d_{c,c'})} \sum_{c'} f(d_{c,c'})
  h_{c,c'}(t) \label{eqn:rate2}
\end{eqnarray}
where $h_{c,c'}(t)$ is given in \eqref{eqn:community-interaction}. The subnetwork interactions are stronger contexts for disease spread. Contact tracing targets exactly these subnetworks for additional testing, case isolation or quarantine.

\subsection{Seeding of infection}

Two methods of seeding the infection have been implemented.

\begin{itemize}
  \item A small number of individuals can be set to either exposed, presymptomatic/asymptomatic, or symptomatic states, at time $t = 0$, to seed the infection. This can be done randomly based either on ward-level probabilities, which could be input to the simulator, or it can be done uniformly at random across all wards in the city.

  \item A seeding file indicates the average number of individuals who should be seeded on each day in the first stage of infectiousness (presymptomatic or asymptomatic). This could be done based on data for patients with a foreign travel history who eventually visited a hospital. A certain multiplication factor then accounts for the asymptomatic and the symptomatic individuals that recover without the need to visit the hospital. The seeding is done at a random time earlier in the time line, based on the disease progression.
\end{itemize}

\subsection{Calibration}
We calibrate our model by tuning the transmission coefficients at various interaction spaces under the no-intervention scenario in order to match the cumulative fatalities to a target curve. We assume a common upscaling factor $\tilde{\beta}$ for the transmission coefficients of smaller subnetworks, i.e., we set $\beta_w^* = \tilde{\beta} \beta_w$, $\beta_s^* = \tilde{\beta}\beta_s$ and $\beta_{h}^* = \beta_{c}^* = \tilde{\beta}\beta_c$. We assume that $\tilde{\beta} = 9$, indicating that the subnetworks account for 90\% of the  overall contacts. The following heuristic iterative algorithm inspired by stochastic approximation is then used to identify the best choice of the free parameters.
\begin{align*}
\beta_h(n+1) &= \left(\beta_h(n) -
\frac{\Lambda_h(n)}{n+3}\right) \times [\exp(m^*- m(n))]^{1/a}_{a}, \\
\beta_w(n+1) &= \left(\beta_w(n) - \frac{\Lambda_w(n)}{n+3}\right) \times [\exp(m^*- m(n))]^{1/a}_{a}, \\
\beta_c(n+1) &= \left(\beta_c(n) - \frac{\Lambda_c(n)}{n+3}\right) \times [\exp(m^* - m(n))]^{1/a}_{a},
\end{align*}
where $[\exp(m^*- m(n))]^{1/a}_{a} = \min\{\max\{\exp(m^*- m(n)), a\}, 1/a\}$, $\Lambda_h(n)$ (resp.~$\Lambda_w(n)$, $\Lambda_c(n)$) is the fraction of infections from home (resp.~workplace, community) in the $n$th step, and $m(n)$ is the slope of the cumulative fatalities curve in log-scale in an initial linear region, obtained by running the simulator on a smaller city file (of 1 million population) in the no intervention scenario. We set $a = 2/3$. To minimise the effect of stochasticity from the simulator, in each step $n$, we run the simulator $6$ times and take the average values of $\Lambda_h(n), \Lambda_w(n), \Lambda_c(n)$ and $m(n)$. We stop the algorithm at time $n$ if we meet our targets, i.e., if
\begin{align*}
|m(n)-m^*| & \leq 0.01,\\
|\Lambda_h(n) - \Lambda_h^*| & \leq 0.01, \\
|\Lambda_w(n) - \Lambda_w^*| & \leq 0.01, \text{ and }\\
|\Lambda_c(n) - \Lambda_c^*| & \leq 0.01,
\end{align*}
where $\Lambda_h^* = \Lambda_w^* = \Lambda_c^* = 1/3$, and $m^*$ is the target slope (the target slope is similarly computed from the cumulative fatalities data in log scale; for example, the India fatalities curve in the range 130-199 gives a slope of $m^* = 0.1803$). Once the slopes are matched, assuming that the simulator starts on 01\,March\,2020, we find the delay between the fatalities curve from the simulator and the target data. We then use the resulting contact rates and the above calibration delay to launch our simulations.

To avoid any oscillatory behaviour of the calibration algorithm, we also set the scale factor in each of the above update steps to be $[\exp((m^*- m(n))/n)]^{1/a}_{a}$ whenever $|\Lambda_h(n) - \Lambda_h^*| \leq 0.02$, $|\Lambda_w(n) -\Lambda_w^*| \leq 0.02$ and $|\Lambda_c(n) - \Lambda_c^*| \leq 0.02$. In addition, we set the scale factor to be $[\exp((m^*- m(n))/(n-25))]^{1/a}_{a}$ if $n \geq 30$.

For the simulation results presented in the case studies, we identified the $m^*$ as follows. We assumed a {\em counterfactual} situation where the national level of infection, up to the date when the lockdown's effect is not yet likely to have been seen, estimated to be 8--10\,April\,2020, is moved to the city under study. There were 199 fatalities in India up to 10\,April\,2020. The India cases and fatalities data is based on daily updates compiled by the European Centre for Disease Prevention and Control \cite{ecdcwebsite}. The counterfactual situation (all of India's infections in the isolated city under study) is to ensure that sufficient data is available for calibration before the national lockdown's effect is encountered. The fatalities up to this date were likely due to contacts prior to the start of the lockdown. The slope of this curve (in the log-domain) gives the $m^*$. The calibration is further done on a smaller version of the city, with 10 lakh population. The resulting parameters are then used on the full-scale city.

We do not calibrate $\beta_T$, the transmission coefficient at transport space. For the calibration step we take this parameter to be zero while tuning the other parameters. A heuristic justification is as follows. Bengaluru travel interactions will likely be captured through the local community interactions, and we keep it zero throughout, even in the case studies. For Mumbai however, local trains are a key mode of daily transportation with a population of the order of 75 lakh travelling daily using this mode in normal times. However, trains were stopped in Mumbai prior to the national lockdown and were running below capacity for at least a week before that. Moreover, the initial infections were seeded by travellers that came from abroad. The primary mode of travel for this group is unlikely to be rail transport. So we disabled the transport space while calibrating by setting $\beta_T = 0$. Subsequently for the trains on/off case study, we used a heuristic calculation of $\beta_T$; see~\cite[Section~IV]{report2}.

The above procedures identify the contact-related parameters. Other parameters are the distance kernel parameters, the parameter $\alpha$ that accounts for crowding in households, the age-stratified interactions, the distribution parameters for individual infectiousness, the probability of severity, etc. These are set as follows:

\begin{table}[ht]
\caption{Model parameters}
\label{tab:params}
\centering
\begin{tabular}{|l|c|c|c|}
 \hline
 Parameter & Symbol & Bengaluru & Mumbai \\
 \hline
 \hline
 Transmission coefficient at home   & $\beta_h$ & 1.0884 (calibrated) & 0.7928 (calibrated)\\
 Transmission coefficient at school & $\beta_s$ & 0.2548 (calibrated) & 0.2834 (calibrated) \\
 Transmission coefficient at workplace & $\beta_w$ & 0.1274 (calibrated) & 0.1417 (calibrated)\\
 Transmission coefficient at community & $\beta_c$ & 0.0169 (calibrated) & 0.0149 (calibrated)\\
 \hline
 Subnetwork upscale factor & $\tilde{\beta}$ & 9 & 9 \\
 Transmission coefficient at transport space & $\beta_t$ & 0 & 0.1506\\
 Household crowding & $1 - \alpha$ & 0.2 & 0.2\\
 Community crowding & $r_c$ & 1 & 2\\
 Distance kernel $f(d) = 1/(1+(d/a)^b)$ & $(a,b)$ & $(10.751,5.384)$ & $(2.709,1.279)$\\
 Infectiousness shape (Gamma distributed) & (shape,scale) & $(0.25, 4)$ & $(0.25, 4)$ \\
 Severity probability & $\Pr\{C_n = 1\}$ & 0.5 & 0.5\\
 Age stratification & $M_{n,n'}$ & Not used & Not used\\
 Project subnetwork size range & $n_{\mathscr{W}(n)}$ & $3-10$ & $3-10$ \\
 Family friends' subnetwork range & no symbol & 2-5 families & 2-5 families \\
 \hline
\end{tabular}
\end{table}

\subsection{Interventions}
The simulator has the capability to accommodate interventions and compliance. Table~\ref{tab:interventions} describes some of the interventions in \cite{ferguson2020report}, some adapted to suit our demographics, and some new interventions involving the nation-wide 40-day `lockdown' in India and various scenarios of `unlock'. These are fairly straightforward to implement -- we modulate an individual's contact rate with an interaction space (both into the interaction space and out of the interaction space) by a suitable factor associated with intervention. For example, one could easily implement and study cyclic exit strategies as done in \cite{karin2020adaptive}. The triggers for cyclic controls could be based on signals such as the number of individuals that are  hospitalised, as done in our soft ward containment. Yet another one is to quarantine or case isolate based on contact tracing, as we will describe next.

\begin{table*}[!ht]
\caption{Interventions.}
\label{tab:interventions}
\begin{center}
\begin{tabular}{||p{0.1\textwidth}|p{0.23\textwidth}|p{0.6\textwidth}||}
 \hline
 \hline
 Label & Policy & Description \\
 [0.5ex]
 \hline
 \hline
 CI & Case isolation at home & Symptomatic individuals stay at home for 7 days, non-household contacts reduced by 90\% during this period, household contacts reduce by 25\%.\\
 \hline
 HQ & Voluntary home quarantine & Once a symptomatic individual has been identified, all members of the household remain at home for 14 days. Non-household contacts reduced by 90\% during this period, household contacts reduce by 25\%. \\
 \hline
 SDO & Social distancing of those aged 65 and over & Non-household contacts reduce by 75\%. \\
 \hline
 LD & Lockdown & Closure of schools and colleges. Only essential workplaces active. For a compliant household, household contact rate doubles, community contact rate reduces by 75\%, workspace contact rate reduces by 75\%. For a non-compliant household, household contact rate increases by 25\%, workspace contact rate reduces by 75\%, and no change to community contact rate.\\
 \hline
 LD40-CI & Lockdown for 40 days & Lockdown for 40 days and then normal activity, but with CI.\\
 \hline
 LD40-PE-CI & One particular phased emergence (PE) from lockdown & Lockdown for 40 days, then CI, HQ and SDO for 14 days. Schools and colleges remain closed during this period. Normal activity resumes after this period with reopening of schools and colleges, but with CI.\\
 \hline
 LD40-PE-SCCI & Another phased emergence from lockdown & Lockdown for 40 days, then CI, HQ and SDO for 14 days. Schools and colleges remain closed during this period (SC). Normal activity resumes after this period but schools and colleges remain closed for another 28 days (SC). CI remains in place throughout.\\
 \hline
 LD40-PEOE-CI & A third type of phased emergence from lockdown & Lockdown for 40 days, then CI, HQ and SDO for 14 days. Schools and colleges remain closed and an odd-even workplace strategy is in place during this period. Normal activity resumes after this period. CI remains in force throughout.\\
 \hline
 \hline
\end{tabular}
\end{center}
\end{table*}

\subsection{Contact tracing}
Our simulator also includes a framework to study the impact of early contact tracing and testing. We assume that contacts of an individual in the smaller networks such as  neighbourhood network, project network, class network and random community network can be identified and tested/quarantined. The current contact tracing protocol quarantines certain primary contacts and tests a subset of these (e.g., symptomatic primary contacts). In our implementation, based on our study of ICMR's testing protocol, given an index case, all household members, a fraction of the friends circle, a fraction of the inner school/workplace circle, and a fraction of the neighbourhood community are termed as primary contacts of this index case. All of these are quarantined, and a fraction of the symptomatic and another fraction of the asymptomatic among these are tested. Those who test positive become new index cases and spawn further contact tracing. The testing fractions are calibrated to match the actual reported cases and the test-positivity rate.

\subsection{Limitations}
We list some limitations of our simulator.
\begin{itemize}
\item We do not have activity modelling in our simulator. As a consequence, weekly and daily patterns on interactions are not taken into account; for instance, the absence of interaction in workplaces and schools during weekends/public holidays, an increased interaction in public transport during morning and evening peak hours etc. are not taken into account in our model. Instead, all these factors are abstracted into a single infection rate for each individual prescribed by~(\ref{eqn:rate}),~(\ref{eqn:rate-age-stratified}) and~(\ref{eqn:rate2}).
\item Some of the data that we need in our simulator, such as the household size distribution, workplace size distribution, school size distribution, commuter distance distribution etc., can perhaps be difficult to obtain for some cities.
\item We have too many free parameters in our model. This can lead to overfitting resulting in high generalisation error.
\item The framework is computationally intensive.
\item Since the disease spread model has quite a bit of stochasticity (e.g., the incubation time), we need to perform multiple runs of our simulator and take an average of the outputs. We do not have an estimate on the variability of our outputs across multiple runs; such an analysis will be essential to determine the number of runs we need to perform in order for our outputs to be close to the average.
\end{itemize}
\section{Algorithmic aspects}
\subsection{Algorithmic aspects related to city generation}
Generation of a synthetic city is performed via the following steps.
\begin{enumerate}
\item \textit{Data gathering and data preparation} involves the following.

    \begin{itemize}

    \item[(a)] Census Data Processing: The primary data sources used to generate the synthetic city are the 2011 decennial census data of India and the intermediate survey reports for a city. The raw data are typically either in spreadsheets or as tables in a PDF document. We use Python packages like pandas and tabula to clean, process and prepare the data required for creating a synthetic city.

        The data required for creating a synthetic city like the ward-wise demographics, employment status, number of households and the origin-destination matrix are created as  separate comma separated values (csv) files. Distributions like the household size, workplace size, school size, for the city are collected as a single JavaScript Object Notation (JSON) file.

        \item[(b)]  \textit{Geo-spatial mapping}: In addition to the census data, the instantiation for the city also requires the geographic representations for a ward like the ward centre and ward boundaries. These are obtained from map files. Map files are mined in different formats like shapefiles (.shp, .shx) or geoJSON (.geojson), which are processed using Python`s geopandas package.
        \end{itemize}

    \item \textit{Instantiating the city files} is done by running a python script on the following inputs: the map data (.geojson), the census data on the demographics(.csv), households(.csv) and employment(.csv) files along with the additional parameters specified in `cityProfile.json'. The instantiation of a synthetic city is done in three stages namely:

    \begin{itemize}
       \item[(a)] \textit{Processing Inputs}: The script ingests the input (.csv) files using the pandas package and computes parameters based on the input like the unemployment fraction, fraction of population in each ward. The GeoJSON file is processed with the geopandas package to parse the input file and the shapely package to compute the ward centre, the ward boundaries and the neighbouring wards. Apart from the data files, the target population for which the instantiation is to be done, the average number of students in a school, the average number of individuals in one workplace are input parameters specified at the start of the script. The age distribution, household-size distribution and school-size distribution are taken as inputs from the `cityProfile.json' file.

        \item[(b)] \textit{Instantiating individuals} comprises of an algorithm that randomly assigns individuals to households by respecting the household-size distribution. Each household has individuals assigned with generational gaps, yet the instantiated population's empirical age distribution must match the given age distribution. Individuals are assigned to workplaces or schools or neither, based on their age and the unemployed fraction, and are assigned the appropriate `workplaceType'. Once an individual is assigned to a household, the location of the individual is mapped to the location of the assigned household. While instantiating households, the ward number to which the household is assigned is specified and based on the ward number the respective ward boundaries are obtained from the map data in the GeoJSON file. The ward boundaries are typically represented as `Polygons'. The location of households , workplaces and schools are randomly sampled as a point inside the ward boundary. For instantiating households in high-density areas, we sample locations either from a GeoJSON file with boundaries of the high-density areas or from a collection of pre-sampled locations of households in high density areas. Common areas where community interactions take place are instantiated at the ward centres, assumed to be the centroids of the polygons. These tasks are accomplished using the following python packages: numpy, random, pandas and shapely. The outputs of this stage are collections of the instantiated individuals, their assigned households, schools, workplaces, transport and community areas.

        \item[(c)] \textit{Additional processing for generating city files}: Before generating the city files, additional processing is done on the dataframes which includes computing the distance of the individuals to their respective ward centres. This stage uses the pandas package for processing and generating the city files in the JSON file format for each instantiated collection namely the individuals, households, workplaces, schools, community centres, and distance between wards.
        \end{itemize}

\end{enumerate}
\subsection{Algorithmic aspects related to disease spread}
The disease progression part of the simulator is broadly implemented as follows. There are four time steps on a given day. At each time step, we go through each susceptible agent and find out the infection rate given by either~(\ref{eqn:rate}),~(\ref{eqn:rate-age-stratified}) or~(\ref{eqn:rate2}) depending on the interaction that we want to model. We then update the disease progression status of that agent based on the infection rate. Once an individual becomes exposed, this person goes through various stages of the disease depending on age, and contributes to the infection rate of other individuals in the individual's interaction space as long as the individual is infectious. Eventually, the agent recovers from the disease or dies. At the end of the simulation, we output a time series of various quantities of interest into a file, such as the daily number of cases, cumulative number of cases, daily number of hospitalisations, daily number of fatalities, cumulative number of fatalities, etc.

Some of the key features of our simulator that help reduce the space and time complexity are as follows.
\begin{itemize}
\item We have a single community space per ward where individuals living in that ward come together and interact. We then have an interaction among communities to model community interactions among people living in different wards. With $n$ agents and $c$ communities, such a model keeps the computational complexity at $O(n) + O(c^2)$. If we had considered a more complex interaction between individuals, where each individual interacts with every other individual living in the city with a certain contact rate, the complexity would have been $O(n^2)$. Modelling of other interaction spaces such as households, schools, workplaces and transport spaces also results in a similar reduction in time complexity from $O(n^2)$ to $O(n)$. Since the number of agents are typically of the order of $10^7$, such a reduction has a huge impact on the running time of the simulator.
\item Contact tracing requires us to maintain a list of contacts made by each agent. In our implementation, we assume that each individual has a certain number of contacts that we can trace (which is random, but independent of $n$). As a result, the space complexity becomes $O(n)$ instead of $O(n^2)$.
\item In the age-stratified interaction as well as OD-matrix based distance kernel, we consider dominant terms of the age-based contact rate matrix as well as OD-matrix by doing a principal component analysis and by focusing on a few important components. This helps simplify the summations in~(\ref{eqn:rate-age-stratified}).
\end{itemize}
These optimisation features appear to be novel features of our simulator.

\section{Conclusion}
In this work, we built an agent-based simulator to study the impact of various non-pharmaceutical interventions in the context of the ongoing COVID-19 pandemic. We demonstrated the capabilities of our simulator via various case studies for Bengaluru and Mumbai. Some of the key features of our simulator include age-stratified interaction that captures heterogeneity in interaction among people in a given interaction space, the ability to implement  various interventions such as soft ward containment, phased opening of workplaces and community spaces, a broad class of contact tracing based testing and case isolation protocols, etc. These features help our simulator to capture the ground reality very well and provide us with realistic predictions. Some future directions include bringing in movement of people into and out of the city and studying the impact of various mobility patterns, modelling and studying the impact on public-health oriented decisions on the economy, incorporating activity modelling into our simulator and using the simulator to obtain district-scale or country-scale predictions. We hope that such agent-based simulators find a regular place in every public health official's tool kit.

\section*{Acknowledgments}

We sincerely thank Arpit Agarwal, Priyanka Agrawal, Bharadwaj Amrutur, M.~S.~Ashwin, Siva Athreya, Abhijit Awadhiya, Chiranjib Bhattacharyya, Vijay Chandru, Avhishek Chatterjee, Arpan Chattopadhyay, Siddhartha Gadgil, Aditya Gopalan, K.~V.~S.~Hari, Ramesh Hariharan, Aniruddha Iyer, Srikanth Iyer, Shipra Jain, Subhasis Jethy, Jacob John, Shruti Kakade, Navin Kashyap, Micky Kedia, DataMeet (a community of Data Science and Open Data enthusiasts), Revanth Krishna, Vrishab Krishna, Manjunath Krishnapur, Anurag Kumar, Rahul Lodhe, Rahul Madhavan, Madhav Marathe, Satyajit Mayor, Gautam Menon, Shiraz Minwalla, Prem Mony, Chandra Murthy, Y.~Narahari, Uma Chandra Mouli Natchu,  Praneeth Netrapalli, Abdul Pinjari,  Vinod Prabhakaran, Dr.~Prabhu, Rishi Prajapati, Ketan Rajawat, G.~Rangarajan, G.~V.~Sagar, Agrim Sharma, Abhishek Sinha, Mukund Thattai, Himanshu Tyagi, Srinivasan Venkataramanan, V.~Vinay, Anil Vullikanti, and many volunteers for their valuable comments and help. 

The SahasraT supercomputing cluster at the Supercomputer Education and Research Centre (SERC), Indian Institute of Science, was used for most of the simulations.

\bibliographystyle{IEEEtran}
{
\bibliography{IEEEabrv,JIISc}
}

\end{document}